\documentclass[twocolumn]{aastex62}

\usepackage{amsmath}

\newcommand{\fermi}{\emph{Fermi}}
\newcommand{\gammaray}{$\gamma$-ray}

\shorttitle{The June 2016 Outburst of the Blazar 3C454.3}
\shortauthors{Weaver et al.}
\accepted{2019 March 9 by ApJ}

\begin{document}

%------------------------------------------------------------------------------------------------------------------------
% 		TITLE AND AUTHORS
%------------------------------------------------------------------------------------------------------------------------

\title{The June 2016 Optical and Gamma-Ray Outburst and Optical Micro-Variability of the Blazar 3C454.3}

\author[0000-0001-6314-0690]{Zachary R. Weaver}
\affiliation{Institute for Astrophysical Research, Boston University, 725 Commonwealth Avenue, Boston, MA 02215, USA}
\affiliation{Colgate University, Department of Physics and Astronomy, 13 Oak Drive, Hamilton, NY 13346, USA}

\author[0000-0002-9844-1730]{Thomas J. Balonek}
\affiliation{Colgate University, Department of Physics and Astronomy, 13 Oak Drive, Hamilton, NY 13346, USA}

\author[0000-0001-6158-1708]{Svetlana G. Jorstad}
\affiliation{Institute for Astrophysical Research, Boston University, 725 Commonwealth Avenue, Boston, MA 02215, USA}
\affiliation{Astronomical Institute, St. Petersburg State University, Universitetskij Pr. 28, Petrodvorets, St. Petersburg 198504, Russia}

\author[0000-0001-7396-3332]{Alan P. Marscher}
\affiliation{Institute for Astrophysical Research, Boston University, 725 Commonwealth Avenue, Boston, MA 02215, USA}

\author[0000-0002-4640-4356]{Valeri M. Larionov}
\affiliation{Astronomical Institute, St. Petersburg State University, Universitetskij Pr. 28, Petrodvorets, St. Petersburg 198504, Russia}
\affiliation{Main (Pulkovo) Astronomical Observatory of RAS, Pulkovskoye shosse 60, St. Petersburg 196149, Russia}

\author{Paul S. Smith}
\affiliation{Steward Observatory, University of Arizona, 933 N. Cherry Ave, Tuscon, AZ 85721, USA}

\author{Samantha J. Boni}
\affiliation{Bridgewater State University, 131 Summer Street, Bridgewater, MA 02324, USA}
\affiliation{Colgate University, Department of Physics and Astronomy, 13 Oak Drive, Hamilton, NY 13346, USA}

\author{George A. Borman}
\affiliation{Crimean Astrophysical Observatory, P/O Nauchny, Crimea 298409, Russia}

\author{K. J. Chapman}
\affiliation{Colgate University, Department of Physics and Astronomy, 13 Oak Drive, Hamilton, NY 13346, USA}

\author{Leah G. Jenks}
\affiliation{Colgate University, Department of Physics and Astronomy, 13 Oak Drive, Hamilton, NY 13346, USA}
\affiliation{Brown University, Department of Physics, 182 Hope Street, Providence, RI 02912, USA}

\author[0000-0001-9518-337X]{Evgenia N. Kopatskaya}
\affiliation{Astronomical Institute, St. Petersburg State University, Universitetskij Pr. 28, Petrodvorets, St. Petersburg 198504, Russia}

\author{Elena G. Larionova}
\affiliation{Astronomical Institute, St. Petersburg State University, Universitetskij Pr. 28, Petrodvorets, St. Petersburg 198504, Russia}

\author[0000-0002-9407-7804]{Daria A. Morozova}
\affiliation{Astronomical Institute, St. Petersburg State University, Universitetskij Pr. 28, Petrodvorets, St. Petersburg 198504, Russia}

\author[0000-0001-9858-4355]{Anna A.  Nikiforova}
\affiliation{Astronomical Institute, St. Petersburg State University, Universitetskij Pr. 28, Petrodvorets, St. Petersburg 198504, Russia}
\affiliation{Main (Pulkovo) Astronomical Observatory of RAS, Pulkovskoye shosse 60, St. Petersburg 196149, Russia}

\author{Alina Sabyr}
\affiliation{Colgate University, Department of Physics and Astronomy, 13 Oak Drive, Hamilton, NY 13346, USA}

\author[0000-0003-4147-3851]{Sergey S. Savchenko}
\affiliation{Astronomical Institute, St. Petersburg State University, Universitetskij Pr. 28, Petrodvorets, St. Petersburg 198504, Russia}

\author{Ryan W. Stahlin}
\affiliation{Colgate University, Department of Physics and Astronomy, 13 Oak Drive, Hamilton, NY 13346, USA}

\author[0000-0002-9907-9876]{Yulia V. Troitskaya}
\affiliation{Astronomical Institute, St. Petersburg State University, Universitetskij Pr. 28, Petrodvorets, St. Petersburg 198504, Russia}

\author[0000-0002-4218-0148]{Ivan S. Troitsky}
\affiliation{Astronomical Institute, St. Petersburg State University, Universitetskij Pr. 28, Petrodvorets, St. Petersburg 198504, Russia}

\author{Saiyang Zhang}
\affiliation{Colgate University, Department of Physics and Astronomy, 13 Oak Drive, Hamilton, NY 13346, USA}

\correspondingauthor{Zachary R. Weaver}
\email{zweaver@bu.edu}

%------------------------------------------------------------------------------------------------------------------------
% 		ABSTRACT AND KEYWORDS
%------------------------------------------------------------------------------------------------------------------------

\begin{abstract}
The quasar 3C454.3 underwent a uniquely-structured multi-frequency outburst in June 2016. The blazar was observed in the optical $R$ band by several ground-based telescopes in photometric and polarimetric modes, at \gammaray\ frequencies by the \fermi\ Large Area Telescope, and at 43 GHz with the Very Long Baseline Array. The maximum flux density was observed on 2016 June 24 at both optical and \gammaray\ frequencies, reaching $S^\mathrm{max}_\mathrm{opt}=18.91\pm0.08$ mJy and $S_\gamma^\mathrm{max} =22.20\pm0.18\times10^{-6}$ ph cm$^{-2}$ s$^{-1}$, respectively. The June 2016 outburst possessed a precipitous decay at both \gammaray\ and optical frequencies, with the source decreasing in flux density by a factor of 4 over a 24-hour period in $R$ band. Intraday variability was observed throughout the outburst, with flux density changes between 1 and 5 mJy over the course of a night. The precipitous decay featured statistically significant quasi-periodic micro-variability oscillations with an amplitude of $\sim 2$-$3\%$ about the mean trend and a characteristic period of 36 minutes. The optical degree of polarization jumped from $\sim3\%$ to nearly 20\% during the outburst, while the position angle varied by $\sim120\degr$. A knot was ejected from the 43 GHz core on 2016 Feb 25, moving at an apparent speed $v_\mathrm{app}=20.3c\pm0.8c$. From the observed minimum timescale of variability $\tau_\mathrm{opt}^\mathrm{min}\approx2$ hr and derived Doppler factor $\delta=22.6$, we find a size of the emission region $r\lesssim2.6\times10^{15}$ cm. If the quasi-periodic micro-variability oscillations are caused by periodic variations of the Doppler factor of emission from a turbulent vortex, we derive a rotational speed of the vortex $\sim0.2c$.
\end{abstract}


%------------------------------------------------------------------------------------------------------------------------
% 		INTRODUCTION
%------------------------------------------------------------------------------------------------------------------------

\section{Introduction}
\label{sec:intro}

\floattable
\begin{deluxetable}{cccccc}
\tablecaption{Telescope Information \label{tab:teleloc}}
\tablecolumns{5}
\tablewidth{0pt}
\tablehead{\colhead{Number} & \colhead{Telescope} & \colhead{Institution} & \colhead{\# Obs.} & \colhead{Lat.} & \colhead{Lon.}}
\startdata
1 & 40-cm Newtonian- & Foggy Bottom Observatory & 135 & $42\degr\ 48\arcmin\ 59\arcsec$ N & $75\degr\ 31\arcmin\ 59\arcsec$ W \\
& Cassegrain& Colgate University, Hamilton, NY, USA & & & \\
2 & 1.83-m Perkins & Lowell Observatory & 60 & $35\degr\ 05\arcmin\ 53\arcsec$ N & $111\degr\ 32\arcmin\ 12\arcsec$ W \\
& & Flagstaff, AZ, USA & & & \\
3 & 70-cm AZT-8 & Crimean Astrophysical Observatory & 139 & $44\degr\ 43\arcmin\ 38\arcsec$ N & $34\degr\ 00\arcmin\ 49\arcsec$ E \\
& & Nauchny, Crimea & & & \\
4 & 40-cm LX-200 & St. Petersburg University & 10 & $59\degr\ 52\arcmin\ 55\arcsec$ N & $29\degr\ 49\arcmin\ 35\arcsec$ E \\
& & St. Petersburg, Russia & & & \\
5$^{*}$ & 1.54-m Kuiper & Steward Observatory &  & $32\degr\ 24\arcmin\ 54\arcsec$ N & $110\degr\ 42\arcmin\ 52\arcsec$ W  \\
 & & Mt. Bigelow, AZ, USA & & & \\
6$^{*}$ &  2.3-m Bok & Steward Observatory & 28 & $31\degr\ 57\arcmin 36\arcsec$ N & $111\degr\ 35\arcmin\ 59\arcsec$ W \\
 & & Kitt Peak, AZ, USA & & & \\
\enddata
\tablecomments{$^*$ Observations from these telescopes were made with the same instrument and are counted together in the number of observations.}
\end{deluxetable}

The blazar 3C454.3 (3FGL J2254.0+1609, $z = 0.859$) is an optically violent, flat spectrum radio quasar noted for being among the brightest \gammaray\ sources in the sky. Since the early 2000s, 3C454.3 has undergone a number of extremely energetic and rapidly variable outbursts across the electromagnetic spectrum \citep[e.g.,][]{Villata2006, Jorstad2010, Vercellone2010, Wehrle2012}. While the basic cause of the extremely high non-thermal luminosity and rapid variability of the flux and polarization in blazars such as 3C454.3 can be explained by a relativistic jet of high-energy plasma  \citep[e.g.,][]{Blandford1979, Marscher1985, Sikora2009}, understanding of the physical processes involved and the mechanism(s) for high energy production remains limited \citep{Jorstad2013}.

During the current age of large-scale surveys, long and concentrated observations of individual objects remain vital due to their ability to provide a wealth of detailed information about an object. Such observations are particularly useful when the source can be observed at many different wavebands \citep{Wehrle2012}, leading to calculations and measurements of timescales of variability, apparent speeds of superluminal knots, and the time-evolution of the spectral energy distribution \citep[e.g.,][]{Marscher2010, Jorstad2010, Britto2016}.

The highest-amplitude optical outburst of 3C454.3 occurred in 2005, with a peak optical brightness of $R=12$ magnitude, triggering a multi-frequency campaign by the Whole Earth Blazar Telescope \citep{Villata2006}. \citet{Villata2007}, \citet{Raiteri2008}, \citet{Hagenthorn2009}, and \citet{Jorstad2010} have analyzed comprehensive multi-frequency observations of this outburst. Several subsequent, smaller outbursts have been intensely studied across the electromagnetic spectrum, including those in 2008 \citep{HagenThorn2013}, 2009 \citep{Raiteri2011}, 2010 \citep{Bachev2011}, and 2014 \citep{Kushwaha2016}.

While 3C454.3 has been observed at optical frequencies as far back as 1899 \citep{Angione1968}, it was first detected at \gammaray\ frequencies in 1992 by EGRET on the \emph{Compton Gamma-Ray Observatory} \citep{Hartman1999}. The blazar was not observed by \gammaray\ telescopes during the outburst of 2005, but has been routinely detected by the \emph{Astro-rivelatore Gamma a Immagini LEggero} (AGILE) \citep{Tavani2009, Vercellone2012} and \emph{Fermi Gamma-Ray Space Telescope} \citep{Abdo2009Early, Atwood2009} orbiting observatories starting in 2007 and 2008, respectively. The blazar was especially bright during a series of \gammaray\ outbursts in late 2009, early 2010, and late 2010 \citep{Ackermann2010, Vercellone2011, Coogan2016}. In November 2010, 3C454.3 reached a flux of $F_\gamma^\mathrm{max} = 8.5 \pm 0.5 \times 10^{-5}$ photons cm$^{-2}$ s$^{-1}$, the highest \gammaray\ flux ever detected from a single, non-transient cosmic source up to that point \citep{Abdo2011}. Analyses of these rich datasets in conjunction with lower-frequency observations serve as a valuable probe into the structure and conditions of the jet within distances of $\lesssim 10$ pc from the central engine \citep{Coogan2016}.

The June 2016 optical and \gammaray\ outburst of 3C454.3 is analyzed in this paper. A time period between 2016 May 1 and 2016 September 30 was selected to concentrate on short-timescale variability using observations from several ground-based telescopes in $R$ band, polarimetric observations, and \gammaray\ observations obtained with the \fermi\ Large Area Telescope (LAT). The observations and data reduction methods are described in $\S$\ref{sec:Obs}. Analyses of the data, including structure and timing of features in the resultant light curves, are given in $\S$\ref{sec:structure}. An investigation into the rapid intraday and micro-variability fluctuations in $R$ band is presented in $\S$\ref{sec:variability}. An analysis of a sequence of Very Long Baseline Array (VLBA) 43 GHz images of 3C454.3 is presented in $\S$\ref{sec:RadioKnots}, yielding a measure of the Doppler factor of an observed radio knot, labeled $K16$. This measurement is used in $\S$\ref{sec:MagFieldStrength} and $\S$\ref{sec:shortvariability}, along with the timescales of variability and observed flux values, to calculate important physical parameters of 3C454.3, including an estimate of the speed of turbulence in the jet. The findings are summarized and concluding remarks are made in $\S$\ref{sec:conclusions}.\\

\begin{figure*}
\plotone{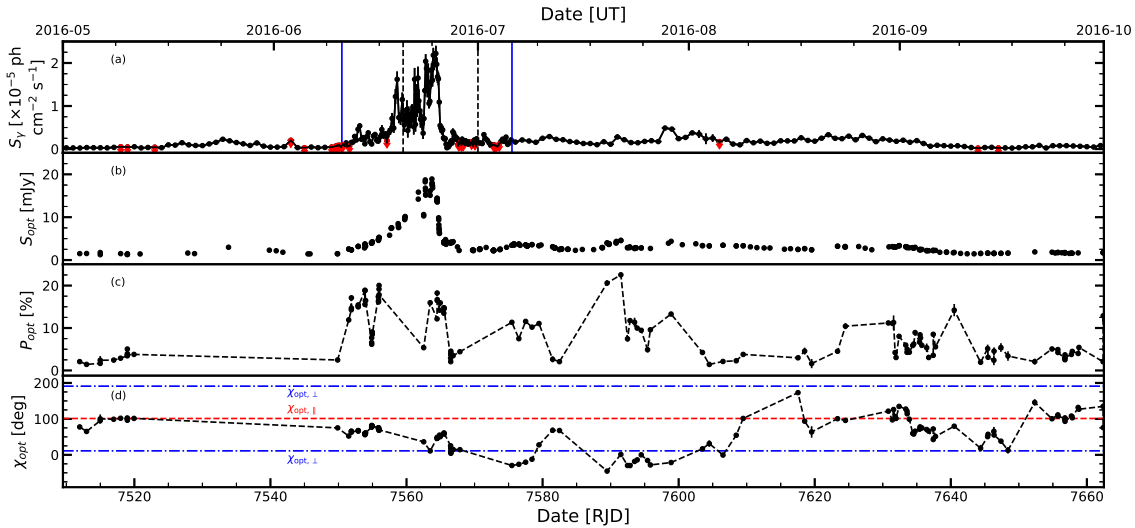}
\caption{Flux and polarization vs. time of 3C454.3. The date 2016 July 1 is RJD: 7570.5. (a) \fermi-LAT \gammaray\ flux with varying time bins; (b) optical light curve in $R$ band; (c) degree of optical linear polarization; (d) position angle ($\chi_\mathrm{opt}$) of optical polarization. In (a), the outer, blue, vertical, solid lines mark the division between one-day and six-hour \gammaray\ binning, while the inner pair of black, vertical, dashed lines mark the division between six-hour and three-hour binning. Upper limits on 24 \fermi-LAT data points are marked with a downward-facing, red arrow. In (d), the horizontal lines correspond to polarization angles that are parallel ($\chi_{\mathrm{opt},\parallel}$, red dashed) and perpendicular ($\chi_{\mathrm{opt},\perp}$, blue dash-dot) to the average parsec-scale jet direction of -$79\degr$ determined using 43 GHz VLBA imaging of the blazar between Jan 2016 and Jun 2017 (see $\S$\ref{sec:RadioKnots}).}
\label{fig:lightcurves}
\end{figure*}

%------------------------------------------------------------------------------------------------------------------------
% 		OBSERVATIONS AND DATA REDUCTION
%------------------------------------------------------------------------------------------------------------------------

\section{Observations and Data Reduction}
\label{sec:Obs}

We analyze data obtained from 2016 May 1 to 2016 September 30 at \gammaray\ energies from 0.1-300 GeV and in the optical Johnson $R$ band. The observations at optical wavelengths, as well as the data reduction at all wavelengths, were performed by the authors. Throughout this paper, dates are referred to using reduced Julian Date, RJD $=$ JD$ - 2,450,000$, as well as the UT date. The analyzed period is RJD: 7509.5 - 7662.5. We adopt current cosmological constants from \citet{Planck2014}: $\Omega_M = 0.308,\ \Omega_\Lambda = 0.692,$ and Hubble Parameter $H_0 = 67.8$ km s$^{-1}$ Mpc$^{-1}$.

\subsection{Multi-frequency Light Curves}
\label{sec:lightcurves}

The \gammaray\ data were collected with the \fermi-LAT. Pass 8 photon and spacecraft data were used, along with version \emph{v10r0p5} of the Fermi Science Tools, the \emph{iso\_P8R2\_SOURCE\_V6\_v06.txt} isotropic template, and the \emph{gll\_iem\_v06} Galactic diffuse emission model.\footnote{Provided at \url{https://fermi.gsfc.nasa.gov/ssc/data/access/lat/BackgroundModels.html}} We use standard analysis cuts of \texttt{evtype = 3} and \texttt{zmax = 90} for the likelihood analysis. Instead of a single time bin for the entire analysis period, the time binning for the flux over different time periods around the peak of the outburst was modified in order to increase the time resolution of the light curve during the highest levels of activity. For the time periods May 1-June 11 (RJD: 7509.5-7550.5) and July 6-September 30 (RJD: 7575.5-7662.5), a time bin of one day was used to ensure a significant detection despite relatively low flux levels. A six-hour time bin was used for the periods June 11-20 (RJD: 7550.5-7559.5) and July 1-6 (RJD: 7570.5-7575.5). Finally, during the time around the peak of the outburst (June 20-July 1, RJD: 7559.5-7570.5), a time bin of three hours was adopted. 

A region of radius $25\degr$ centered on 3C454.3 was chosen for this analysis. The \gammaray-emission from 3C454.3 and other point sources within a 15$\degr$ radius region of interest of the blazar were represented with spectral models as found in the 3FGL catalog of sources detected by the LAT \citep{Acero2015}, creating a standard annulus with a thickness of $10\degr$ around the region of interest. Specifically, the energy spectrum of 3C454.3 was modeled as a power-law with an exponential cut-off \citep[see][]{Acero2015} of the form

\begin{equation*}
\frac{\mathrm{d}N}{\mathrm{d}E} =  N_0 \left( \frac{E}{E_0} \right)^{\gamma_1} \exp{\left(- \frac{E}{E_\mathrm{c}} \right)^{\gamma_2}\ .}
\end{equation*}

\noindent During the unbinned likelihood analysis\footnote{As described in the \fermi\ Analysis Threads: \url{https://fermi.gsfc.nasa.gov/ssc/data/analysis/scitools/python_tutorial.html}} to compute the light curve, the spectral parameters of all sources within the region of interest were kept fixed to the values listed in the 3FGL, as were the spectral parameters of the quasar, which are $E_0 = 412.75$ MeV for the scale factor, $E_\mathrm{c} = 25.65$ MeV for the cutoff energy, and $\gamma_1 = 1.63$ and $\gamma_2 = 0.28$ for the two power-law indices (Acero et al. 2015). The prefactor $N_0$ was allowed to vary for 3C454.3, as were the normalization parameters of the isotropic emission template and Galactic diffuse emission model. This procedure produces a \gammaray\ light curve with 275 measurements of 3C454.3. The source was considered detected if the test statistic, TS, provided by the maximum-likelihood analysis exceeded 10, which corresponds to approximately a $3\sigma$ detection level \citep{Nolan2012}. Upper-limits were calculated using the standard procedure for the 24 data points with TS $<10$.

The optical photometric data in $R$ band were collected at various telescopes listed in Table~\ref{tab:teleloc}. The data reduction of observations from the Perkins, CAO, and St. Petersburg telescopes is described in \citet{Larionov2008} and \citet{Jorstad2010}. The reduction of the Steward Observatory data is described in \citet{Smith2009}. The observations from the Colgate University Foggy Bottom Observatory (FBO) are described in more detail below.

Between 2016 May 1 and September 30, 860 images of 3C454.3 on 41 nights were taken at FBO with \emph{Photometrics Star 1} CCD systems. The images are primarily two-minute exposures with the $R$ filter designed by \citet{Beckert1989} to conform to the Johnson-Cousins system (central wavelength $\lambda_c = 640$ nm, bandwidth $\Delta \lambda = 160$ nm, with magnitude-to-flux conversion coefficient for a 15 magnitude star $C_\lambda = 3.08$ mJy). The data were reduced using standard IRAF\footnote{IRAF is distributed by the National Optical Astronomy Observatory, which is operated by the Association of Universities for Research in Astronomy, Inc., under cooperative agreement with the National Science Foundation.} V2.12 packages and customized scripts written to facilitate the data handling. 

The images were processed using aperture photometry with the IRAF \emph{apphot} package using a 10\arcsec\ diameter aperture and a sky annulus of inner diameter 24\arcsec\ and outer diameter 44\arcsec. Star 8 of \citet{Smith1998}, with a known magnitude of $R = 13.10 \pm 0.03$, was used as the primary comparison star. A faint star $\sim 10$\arcsec\ west of star 8 contributes $<1\%$ to the brightness of star 8, and is ignored. All error bars presented in this paper are $1\sigma$ uncertainty and were calculated using \emph{apphot}. The validity of these errors was verified by measuring the scatter within a night, as well as over the entire data set for several comparison stars of different magnitudes.
All data points of 3C454.3, unless otherwise stated in the captions to figures, represent the average of 12 images if the magnitude of the blazar $R > 15$, or the average of 6 images if $R < 15$. This was done in an attempt to increase the temporal resolution of data obtained at FBO at high flux levels. The range of $R$ binned in this way was $15.84 \geq R \geq 13.03$ (a flux range of $1.42 \leq S_\mathrm{opt} \leq 18.91$ mJy). Discrepancies between the flux scales of data from the different telescopes is $\sim$0.05 magnitude. However, analysis of data taken at a similar time from telescopes with different reduction methods indicate that no systematic offset is necessary, so no correction is made. No interstellar extinction was factored into the conversion between magnitude and flux. The main comparison stars used have similar color indices as that of the blazar, so we make no correction for atmospheric absorption.

Figure~\ref{fig:lightcurves} shows the \gammaray\ and optical light curves from 2016 May 1 to September 30 (RJD: 7509.5-7662.5) in panels (a) and (b), respectively. Upper limits are denoted by red downward facing arrows in panel (a). Visual inspection of the light curves reveals a $\sim3$ week period of high activity at both \gammaray\ and optical frequencies, from June 11 to July 1 (RJD: 7550.5-7570.5), with lower-level activity occurring both before and after.

\begin{figure*}
\plotone{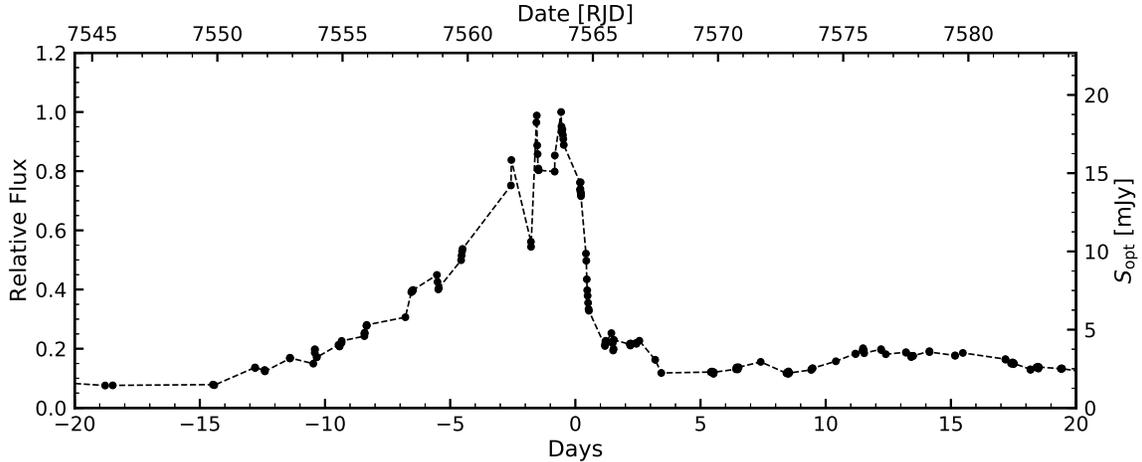}
\caption{The observed optical $R$ band light curve of the June 2016 outburst, relative to $S^{\mathrm{max}}_{\mathrm{opt}} = 18.91 \pm 0.08$ mJy and centered on $T^{\mathrm{max}}_{\mathrm{\gamma}} = 7564.3$ (see Table~\ref{tab:flareparam}). Error bars are included, but in many cases are smaller than the symbols.}
\label{fig:optcurve}
\end{figure*}

\subsection{Observations of Polarization}
\label{sec:PolObs}

Optical linear polarization measurements were performed at telescopes 2-6. The polarization measurements made using telescopes 2 and 3 were obtained in $R$ band, while those from telescope 4 were in white light. The measurements made with telescopes 5 and 6 were spectropolarimetric observations spanning the spectral range 4000-7550 \AA\ at a resolution of 15-20 \AA. Here, we report the polarization averaged over the range of 5000 to 7000 \AA. The details of optical polarization observations and data reduction for these telescopes can be found in \citet{Larionov2008} and \citet{Jorstad2010}. The spectropolarimetric observations of 3C454.3 at Steward Observatory were obtained as part of a program to monitor bright \gammaray\ blazars from the \fermi\ blazar list during the first 10 years of the \fermi\ mission.\footnote{\url{http://james.as.arizona.edu/~psmith/Fermi}} Details of the spectropolarimetric data reduction can be found in \citet{Smith2009}, and the results are discussed in $\S$\ref{sec:opticalpolarization}. The combined optical polarization data obtained from the telescopes used for this study consist of 128 measurements of the degree, $P_\mathrm{opt}$, and electric-vector position angle, $\chi_\mathrm{opt}$. 

The degree of polarization from the source (Fig.~\ref{fig:lightcurves}c) increased from a few percent to $\sim 20\%$ at roughly the same time as the beginning of the outburst at optical and \gammaray\ frequencies, near mid-June 2016. The value of $P_\mathrm{opt}$ during the outburst and the post-outburst period was quite variable, ranging between the pre-outburst level to the maximum. The value of $P_\mathrm{opt}$ decreased over a longer timescale than the fluxes at optical and \gammaray\ frequencies. Over the course of the main outburst, flare $a$ (defined visually as the period of the outburst with the highest $\gamma$-ray flux, from 2016 June 19 to June 25; RJD 7558-7564 -- see $\S$\ref{sec:gammaray}), $\chi_\mathrm{opt}$ varied over a range $\sim 120\degr$, remaining stable at times for only several days (Fig.~\ref{fig:lightcurves}d). During the peak of the outburst, $\chi_\mathrm{opt}$ varied erratically.

%------------------------------------------------------------------------------------------------------------------------
% 		LIGHTCURVE STRUCTURE
%------------------------------------------------------------------------------------------------------------------------

\section{Structure and Timescales of the Outburst}
\label{sec:structure}

\subsection{Optical Outburst}
\label{sec:opticaloutburst}

%Big Table

\floattable
\begin{deluxetable}{lc|lc}
\tablecaption{Parameters of the 2016 Outburst \label{tab:flareparam}}
\tablecolumns{5}
\tablewidth{0pt}
\tablehead{\colhead{\gammaray\ Parameter} & \colhead{Value} & \colhead{Optical Parameter} & \colhead{Value}}
\startdata
$M$ & 275 (24) & $M$ & 120 \\
$\Delta T_\gamma$ [days] & 86 & $\Delta T_\mathrm{opt}$ [days] & 20 \\
$\langle S_\gamma \rangle$ [10$^{-6}$ photons cm$^{-2}$ s$^{-1}$] & 5.44 $\pm$ 5.76 & $\langle S_\mathrm{opt} \rangle$ [mJy] & 7.21 $\pm$ 5.15 \\
$\langle \sigma_\gamma \rangle$ [10$^{-6}$ photons cm$^{-2}$ s$^{-1}$] & 0.96 &  $\langle \sigma_\mathrm{opt} \rangle$ [mJy] & 0.07  \\
\hline
$\Delta T^{a}_\gamma$ [days] & 6.3 & $\Delta T^\mathrm{peak}_\mathrm{opt}$ [days] & 5.82 \\
$T^{\mathrm{max}}_\gamma$ & $\sim$ 2016 Jun 24 19:00 & $T^\mathrm{max}_\mathrm{opt}$ & 2016 Jun 24 05:30  \\
$T^{\mathrm{max}}_\gamma$ [RJD] & 7564.3 & $T^\mathrm{max}_\mathrm{opt}$ [RJD] & 7563.731 \\
$S^{\mathrm{max}}_\gamma$ [10$^{-6}$ photons cm$^{-2}$ s$^{-1}$] & 22.20 $\pm$ 0.18 & $S^\mathrm{max}_\mathrm{opt}$ [mJy] & 18.91 $\pm$ 0.08  \\
$\tau^a_\gamma$ [hr] & 2.63 & $\tau^\mathrm{min}_\mathrm{opt}$ [hr] & 1.97 \\
$f_\gamma^a$ & 3.13 & $f_\mathrm{opt}$ & 1.05 \\
\hline
$\Delta T^{\mathrm{pre}}_\gamma$ [days] & 6 &  $\dots$ & $\dots$ \\
$S^\mathrm{pre}_\gamma$ [10$^{-6}$ photons cm$^{-2}$ s$^{-1}$] & 2.75 $\pm$ 1.18 &  $\dots$ & $\dots$ \\
$\tau^\mathrm{pre}_\gamma$ [hr] & 6.33 &  $\dots$ & $\dots$ \\
$f_\gamma^\mathrm{pre}$ & 2.58 &  $\dots$ & $\dots$ \\
\hline
$\Delta T^{\mathrm{post}}_\gamma$ [days] & 7.5 &  $\dots$ & $\dots$ \\
$S^\mathrm{post}_\gamma$ [10$^{-6}$ photons cm$^{-2}$ s$^{-1}$] & 1.63 $\pm$ 0.74 &  $\dots$ & $\dots$ \\
$\tau^\mathrm{post}_\gamma$ [hr] & 2.86 &  $\dots$ & $\dots$ \\
$f_\gamma^\mathrm{post}$ & 2.86 &  $\dots$ & $\dots$ \\
\hline
$T^b_\gamma$ & 2016 Jul 28 &  $\dots$ & $\dots$ \\
$T^b_\gamma$ [RJD] & 7598.0 &  $\dots$ & $\dots$ \\
$S^b_\gamma$ [10$^{-6}$ photons cm$^{-2}$ s$^{-1}$] & 4.88 $\pm$ 0.32 &  $\dots$ & $\dots$ \\
\hline
$T^c_\gamma$ & 2016 Aug 23 &  $\dots$ & $\dots$ \\
$T^c_\gamma$ [RJD] & 7623.6 &  $\dots$ & $\dots$ \\
$S^c_\gamma$ [10$^{-6}$ photons cm$^{-2}$ s$^{-1}$] & 3.32 $\pm$ 0.32 &  $\dots$ & $\dots$ \\
\hline
$\tau^\mathrm{min}_\gamma$ [hr] & 2.63 &  $\tau^\mathrm{min}_\mathrm{opt}$ [hr] & 1.97 \\
$ T^{\tau_\mathrm{min}}_\gamma$ [RJD] & 7561.06 & $T^{\tau_\mathrm{min}}_\mathrm{opt}$ [RJD] & 7564.512 \\
$ \langle \tau_{\gamma,2} \rangle$ [hr] & 34 $\pm$ 20 & $\langle \tau_{\mathrm{opt},2}  \rangle$ [hr] & 38 $\pm$ 14 \\
\enddata
\tablecomments{\textbf{$\bf\gamma$-\textbf{ray}\ Parameters:} $M$: Number of observations (number of upper-limits);
$\Delta T_\gamma$: Duration of the \gammaray\ outburst; 
$\langle S_\gamma \rangle$: The average flux during the outburst, and its standard deviation; 
$\langle \sigma_\gamma \rangle$: The average 1$\sigma$ uncertainty of an individual measurement during the outburst; 
$\Delta T^{a}_\gamma$: Duration of the main flare (FWHM, see text Section~\ref{sec:gammaray});
$T^{\mathrm{max}}_\gamma$: The date of maximum of the \gammaray\ outburst; 
$S^{\mathrm{max}}_\gamma$: Flux at the peak of the \gammaray\ outburst over a 3 hour bin;
$\tau^a_\gamma$: Minimum timescale of variability of $\gamma$-ray flux during the main flare;
$f^a_\gamma$: Factor of the $\gamma$-ray flux change over $\tau^a_\gamma$;
$\Delta T^\mathrm{pre}_\gamma$: Duration of the pre-flare plateau for flare $a$;
$S^\mathrm{pre}_\gamma$: The average \gammaray\ flux and its standard deviation over period of $\Delta T^\mathrm{pre}_\gamma$;
$\tau^\mathrm{pre}_\gamma$: Minimum timescale of variability of \gammaray\ flux during  $\Delta T^\mathrm{pre}_\gamma$;
$f^\mathrm{pre}_\gamma$: Factor of the \gammaray\ flux change over $\tau^\mathrm{pre}_\gamma$;
$\Delta T^\mathrm{post}_\gamma, S^\mathrm{post}_\gamma, \tau^\mathrm{post}_\gamma, f^\mathrm{post}_\gamma$: Parameters for the post-flare plateau obtained in the same manner as for the pre-flare plateau;
$T^b_\gamma, S^b_\gamma$: Epoch and maximum flux for flare $b$, calculated in the same manner as flare $a$;
$T^c_\gamma, S^c_\gamma$: Epoch and maximum flux for flare $c$, calculated in the same manner as flare $a$;
$\tau^{\mathrm{min}}_\gamma$ (hr): Minimum timescale of variability of \gammaray\ flux during an outburst;
$T^{\tau_{\mathrm{min}}}_\gamma$: Epoch of the start of an event with minimum timescale of variability;
$\langle \tau_{\gamma,2} \rangle$: Typical timescale of flux doubling (see text).
\textbf{Optical Parameters:} $M$: Number of observations;
$\Delta T_\mathrm{opt}$: Duration of optical outburst;
$\langle S_\mathrm{opt} \rangle$: Average flux-density during the outburst and its standard deviation;
$\langle \sigma_\mathrm{opt} \rangle$: The average 1$\sigma$ uncertainty of an individual measurement during the outburst;
$\Delta T^\mathrm{peak}_\mathrm{opt}$: Duration of the main flare (FWHM);
$T^\mathrm{max}_\mathrm{opt}$: Epoch of maximum during the optical outburst;
$S^\mathrm{max}_\mathrm{opt}$: Maximum flux-density and error of the optical outburst;
$\tau^\mathrm{min}_\mathrm{opt}$: Minimum timescale of variability during the optical outburst; 
$f_\mathrm{opt}$: Factor of the flux change over $\tau^\mathrm{min}_\mathrm{opt}$;
$T^{\tau_\mathrm{min}}_\mathrm{opt}$: Epoch of start of an event with minimum timescale of variability;
$\langle \tau_{\mathrm{opt},2} \rangle$: Typical timescale of flux doubling (see text).}
\end{deluxetable}

A detailed subset of the $R$ band optical light curve of the outburst is presented in Figure~\ref{fig:optcurve}. The light curve is normalized to the maximum flux density of the outburst, $S_\mathrm{opt}^\mathrm{max} = 18.91 \pm 0.08$ mJy, and centered on the date of maximum \gammaray\ flux ($T_\gamma^\mathrm{max} \approx$ June 24 19:00, RJD: 7564.3) with a range of $\pm\ 20$ days. Characteristic parameters for the optical light curve can be found in Table~\ref{tab:flareparam}.

The maximum flux density value $S_\mathrm{opt}^\mathrm{max}$ occurred on $T_\mathrm{opt}^\mathrm{max} = $ June 24, 05:30 (RJD: 7563.7314), slightly before the \gammaray\ maximum. The time of maximum flux occurred late into the outburst; the rising time for the optical outburst was $\sim 14$ days, and the decay occurred on a timescale $< 5$ days. The flare profile has a skewed shape that is also apparent in the \gammaray\ outburst (see Figure~\ref{fig:gammaonly} and $\S$\ref{sec:gammaray}), and differs from past optical outbursts \citep[see, e.g.,][]{Ogle2011, Jorstad2013, Kushwaha2016}. Also, unlike the optical outbursts analyzed in \citet{Jorstad2013}, there are no evident pre- and post-outburst plateaus in the optical light curve. We use the full width at half-maximum (FWHM) of a Gaussian function that fits the flare profile near the maximum flux density to define the duration of the optical outburst despite the asymmetry of the light curve, $\Delta T_\mathrm{opt}^\mathrm{peak} \sim 5.8$ days.

We determine timescales of optical flux variability ($\tau_\mathrm{opt}$) using a formalism suggested by \citet{Burbidge1974} and utilized by \citet{Jorstad2013}: $\tau = \Delta t / \ln{(S_2 / S_1)}$, where $S_i$ is the flux density at epoch $t_i$, with $S_2 > S_1$, and $\Delta t = |t_2 - t_1|$. The timescale of variability was calculated for all possible pairs of flux measurements within 3 days of each other if, for a given pair, $S_2 - S_1 > \frac{3}{2} (\sigma_{S_1} + \sigma_{S_2})$, where $\sigma_{S_i}$ is the uncertainty for an individual measurement. The minimum timescale of variability, $\tau_\mathrm{opt}^\mathrm{min}$, is very short: $\sim$2.0 hr, indicative of intraday and perhaps micro-variability (see $\S$\ref{sec:variability}). Such episodes of extreme variability occur infrequently in 3C454.3, with the majority of the active periods exhibiting a timescale of flux doubling between 1 and 2 days. 

A remarkable feature of the optical light curve is the precipitous decay in flux density over one day on June 25 (RJD: 7564.5). The flux density in $R$ band decreased by a factor of $\sim 4$, from 14.2 mJy to 3.8 mJy. Due to the sampling rate of the light curve, the observed timescale of 24 hr is likely an overestimation, as the bulk of the decay occurred over a 4.5 hr time period when the flux density decreased by $\sim2$, from 11.5 to 6 mJy. An in-depth discussion of the optical variability is given in $\S$\ref{sec:variability}. \\

\begin{figure}
\plotone{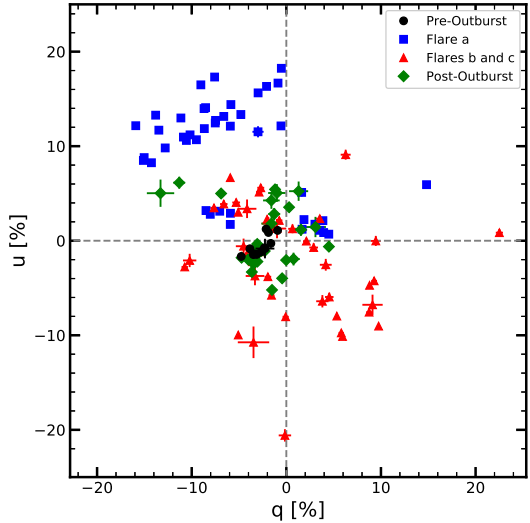}
\caption{Stokes $q$ and $u$ changes over the course of the 2016 outburst. The time ranges are defined following the \gammaray\ outburst (see $\S$\ref{sec:gammaray}): Pre-Outburst - from 2016 May 01 to June 12 (RJD: 7509-7551); Flare a - from June 12 to July 4 (RJD: 7551-7573); Flare b and c - from July 4 to September 6 (RJD: 7573-7637); and Post-Outburst - from September 6 to October 4 (RJD: 7637-7665). \label{fig:QUStokes}}
\end{figure}

\subsection{Optical Polarization}
\label{sec:opticalpolarization}

While the sampling of the optical polarization during the outburst was significantly less intense than that of the optical flux density, there are a few noteworthy features of the polarization curves presented in Figure~\ref{fig:lightcurves}. Prior to the outburst, the degree of optical polarization was low ($P_\mathrm{opt} = 2.47 \pm 0.39 \%$ during May). However, the large gap in sampling between May and June makes it difficult to accurately describe the nature of the polarization prior to the outburst. During the outburst the average polarization $\langle P_\mathrm{opt} \rangle = 12.77 \pm 5.47\%$, with a maximum value of $P_\mathrm{opt}^\mathrm{max} = 20.03 \pm 0.10 \%$ on June 16 (RJD: 7555.97). This is displaced by 8 days from the peaks of the optical and \gammaray\ light curves on Jun 24 (RJD: 7564.3). The quasar 3C454.3 was highly polarized during the June 2016 outburst, while the blazar was in a weakly-polarized state prior to the outburst. The polarization at the peak of the \gammaray\ and optical light curves was $P_\mathrm{opt} = 12.16 \pm 0.15\%$, near the average for the outburst. While both the \gammaray\ and optical outbursts experienced dramatic and precipitous decays on June 25 (RJD: 7564.5), a decrease in the polarization was not seen until June 26 (RJD: 7565.5), when $P_\mathrm{opt}$ fell from $14.55 \pm 0.48 \%$ to $2.91 \pm 0.33\%$ over a 24-hour time period. The value of $P_\mathrm{opt}$ later increased up to $>20\%$, a higher level than during the outburst, for a short period of time, before settling down to near pre-outburst levels. According to the available data obtained prior to and after the outburst, changes in the degree of polarization were more chaotic after than prior to the outburst (despite the sparser sampling in May 2016).

One month prior to the start of the outburst, the emission from 3C454.3 was polarized roughly parallel to the jet direction (see $\S$\ref{sec:RadioKnots}). Throughout the course of the outburst, $\chi_\mathrm{opt}$ rotated in an irregular fashion by $\sim 120\degr$. This change took place over a longer timescale than the rise times of either the optical or \gammaray\ outbursts. It is interesting to note that the changes in $P_\mathrm{opt}$ and $\chi_\mathrm{opt}$ do not coincide. The value of $\chi_\mathrm{opt}$ did not return to a direction nearly parallel to the jet until flare $c$ (described in the following section). Similar high-amplitude drifts in $\chi_\mathrm{opt}$ in 3C454.3 have been noted previously \citep[e.g.,][]{Jorstad2007}. 

While $P_\mathrm{opt}$ was chaotic over the course of the outburst, and $\chi_\mathrm{opt}$ rotated in an irregular fashion by $\sim120\degr$, the polarized light from 3C454.3 followed a general trend over the course of the outburst. Prior to the outburst, the average polarization was low, with average Stokes parameters $\langle q_\text{pre} \rangle =  -2.70 \pm 1.06\%$ and $\langle u_\text{pre} \rangle = -0.55 \pm 1.05\%$, where $1.06\%$ and $1.05\%$ represent the standard deviation of the average values, while the typical uncertainty on a measurement of $q$ or $u$ is $\langle \sigma \rangle = 0.36\%$. During the main part of the outburst, \gammaray\ flare $a$ (see $\S$\ref{sec:gammaray}), $q$ and $u$ became much higher as well as more random than the pre-outburst state, with $\langle q_a \rangle = -5.58 \pm 6.63\%$ and $\langle u_a \rangle = 9.45 \pm 5.41\%$ (see Figure~\ref{fig:QUStokes}). Although there is clustering of $u$ at high values during Flare $a$, several low values are measured as well. This behavior can be connected with the complex structure of the flare seen in the $R$ band light curve (Fig.~\ref{fig:optcurve}). However, sparser sampling of the polarization data does not allow us to investigate a detailed correlation between the degree of polarization and flux behavior. As the outburst faded through flares $b$ and $c$, $q$ and $u$ became more erratic around a central low polarization, with $\langle q_{bc} \rangle = 0.35 \pm 6.36\%$ and $\langle u_{bc}\rangle = -2.04 \pm 5.66\%$. After the outburst, $\langle q_\text{post} \rangle = -2.30 \pm 3.78\%$ and $\langle u_\text{post} \rangle = 0.61 \pm 3.35\%$. The increase in the standard deviation of $\langle q \rangle$ and $\langle u \rangle$ indicates that the post-outburst state was more turbulent than the pre-outburst state. The high-amplitude fluctuations of $q$ and $u$ early in the outburst and the clustering of measurements around low polarizations near the end of the outburst is consistent with the interpretation that the jet contains a superposition of ordered and turbulent magnetic fields \citep{Marscher2017}.

Spectropolarimetric measurements were obtained in addition to the photometric measurements using the same instrument on telescopes 5 and 6. While a complete analysis of the spectra is beyond the scope of this work, we briefly describe the general trends. In order to describe these trends, we rotate the $q$ and $u$ Stokes spectra so that $u^\prime$ averages to $0$ between $5000$ and $7000$ \AA. In this frame, nearly all of the polarization is given by $q^\prime$, and we can avoid the complications of the statistical bias in measurements of $P_\mathrm{opt}$ arising from that parameter's non-normal error distribution \citep{Wardle1974}. For each spectrum of $q^\prime$ and $u^\prime$, the median values in two widely separated wavelength bins, each 500 \AA\ wide, were taken to analyze $P_\mathrm{opt} (\lambda)$. A blue region centered on $\lambda = 4750$ \AA\ and a red region centered on $\lambda = 7250$ \AA\ were chosen to avoid any major emission line features. We then construct $\Delta q^\prime = q^\prime (\text{red}) - q^\prime (\text{blue})$ and $\Delta u^\prime = u^\prime (\text{red}) - u^\prime (\text{blue})$. In this representation, the wavelength dependence of $P_\mathrm{opt}$ is quantitatively given by $\Delta q^\prime$, while $\Delta u^\prime$ indicates the strength of the dependence of $\chi_\mathrm{opt}$ on wavelength.

\begin{figure}
\plotone{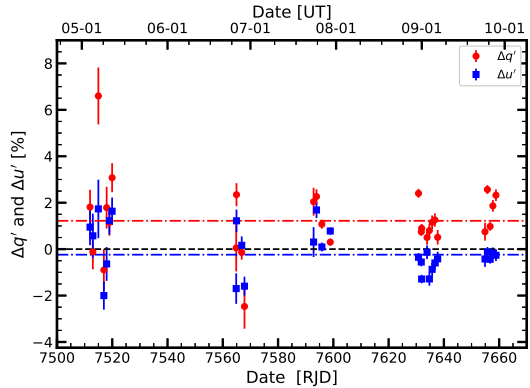}
\caption{The rotated differential Stokes parameters $\Delta q^\prime$ (red circles) and $\Delta u^\prime$ (blue squares) vs. time for the 2016 outburst of 3C454.3. The dot-dash lines show the weighted average of the values. The black dashed line at 0 is included for comparison. The uncertainties are derived from the value $\text{RMS}/\sqrt{N}$ for each region ($\lambda = 7000$-$7500$\ \AA\ (red) and $\lambda = 4000$-$4500$\ \AA\ (blue)), where $N$ is the number of pixels in the spectral region (126 for both regions) and RMS is the root-mean-square calculated from the pixels within the sample region.\label{fig:DQDU}}
\end{figure}

\begin{figure}
\begin{center}
\includegraphics[width=0.4\textwidth, trim=100 200 100 225]{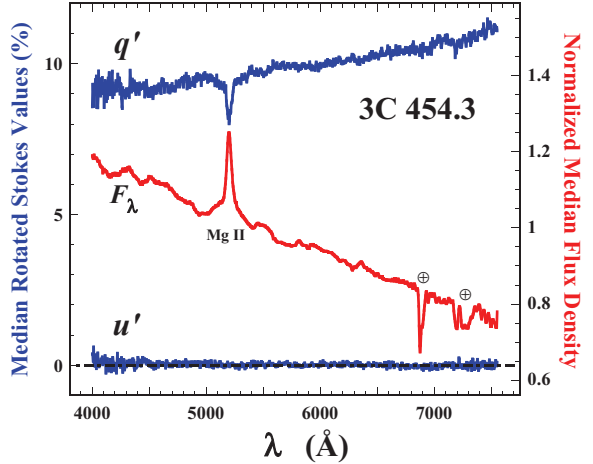}
\end{center}
\caption{Median $q^\prime$ and $u^\prime$ (rotated $q$ and $u$ such that $\langle u^\prime \rangle = 0$ for the wavelength range $5000$-$7000$ \AA) of 3C454.3 for 758 spectropolarimetric observations obtained from 2008-2018 in blue. See text for details. The median optical flux spectrum of 3C454.3 over the same period is shown in red. The symbol $\oplus$ indicates absorption features due to the atmosphere.\label{fig:QWaveDep}}
\end{figure}

During the outburst, the weighted average and propagated uncertainty $\langle \Delta u^\prime \rangle = -0.24 \pm 0.05\%$ (with a standard deviation of $1.01\%$, thus a roughly equal distribution around $\Delta u^\prime = 0$), indicating that $\chi_\mathrm{opt}$ is not strongly dependent on wavelength. However, $\Delta q^\prime$ shows a bias towards positive values, with $\langle \Delta q^\prime \rangle = 1.22 \pm 0.05\%$ and a standard deviation of $1.54\%$. This bias is likely due to the unpolarized blue bump emission diluting the polarization from the non-thermal emission in the jet \citep{Smith1988}, and shows that $P_\text{opt}$ increases with increasing wavelength. The values of $\Delta q^\prime$ and $\Delta u^\prime$ are shown in Figure~\ref{fig:DQDU}, along with their weighted averages. The majority of the spectropolarimetry data were obtained before and after the main outburst, flare $a$, while the optical emission of the blazar was weak ($<5$ mJy) and the polarization only moderately high ($\sim 10\%$). As a result, the observed increase of $P_\text{opt}$ with increasing wavelength supports the findings of \citet{Jorstad2013}, who found a similar trend for weak emission/moderate polarization states of 3C454.3. These trends are seen not just in this outburst, but also in the 10-year spectropolarimetric monitoring of 3C454.3, with $\langle \Delta q^\prime_\mathrm{tot} \rangle = 0.827 \pm 0.001\%$ and $\langle \Delta u^\prime_\mathrm{tot} \rangle = -0.063 \pm 0.008\%$, and standard deviations of $1.310\%$ and $1.002\%$, respectively.

Figure~\ref{fig:QWaveDep} shows the median $q^\prime$ and $u^\prime$ spectra for all Steward Observatory observations from 2008-2018. As expected from the adopted rotation of the Stokes parameters, $u^\prime \approx 0$ across the spectrum, but $q^\prime$ shows a general decrease towards the blue. In the figure, each $q^\prime$ spectrum was normalized such that the median value in the $5500$-$6500$ \AA\ region was set to 10\%. Fig.~\ref{fig:QWaveDep} also shows the median optical flux spectrum of 3C454.3, with the average flux density in the $5400$-$5600$\ \AA\ range normalized to 1. The decrease in $q^\prime$ at the wavelengths corresponding to the Mg II line emission indicates that the broad-line region is unlikely to have a strong polarization.

\begin{figure*}
\plotone{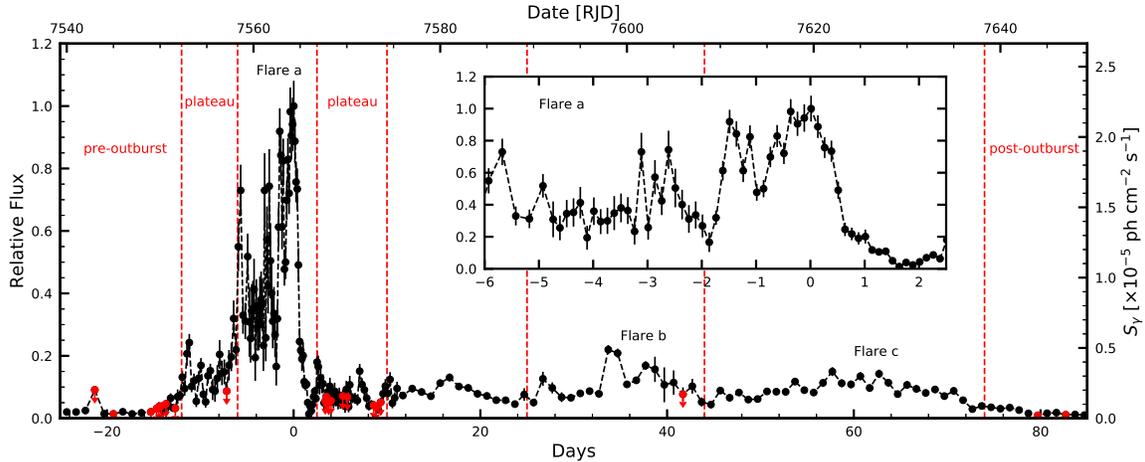}
\caption{Gamma-ray light curve of the June 2016 outburst, relative to $T^{\mathrm{max}}_\gamma = 7564.3$ and normalized to $S^{\mathrm{max}}_\gamma = 22.20 \pm 0.18 \times 10^{-5}$ photons cm$^{-2}$ s$^{-1}$ (see Table~\ref{tab:flareparam}). Upper limits are denoted with red downward arrows. Flares $a$, $b$, and $c$, pre- and post-flare $a$ plateaus, and pre- and post-outburst times are marked with dotted lines. The inset figure shows the shape of the flare $a$ in more detail in the units of the main figure. Flares $b$ and $c$, while low-amplitude, are comparable to flares $b$ and $c$ presented in \citet{Jorstad2013}.}
\label{fig:gammaonly}
\end{figure*}

\begin{figure*}
\plotone{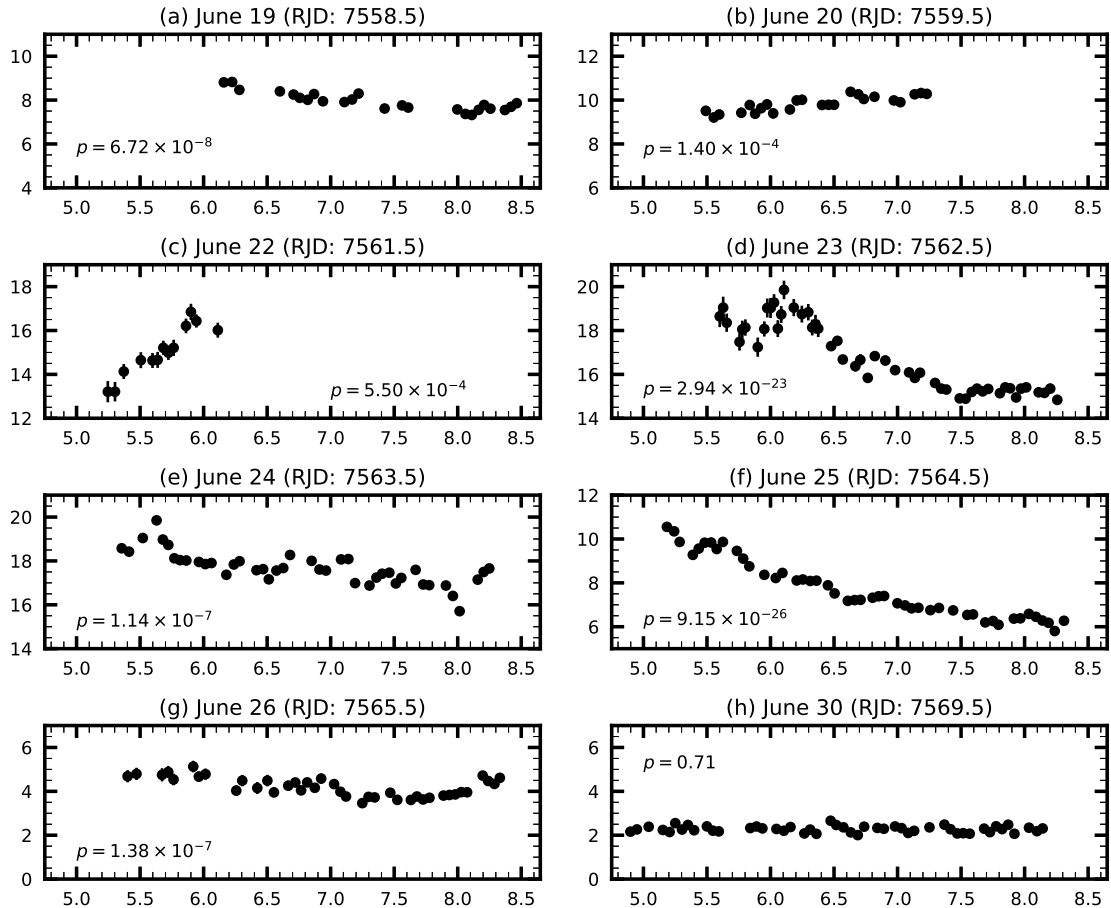}
\caption{Flux density vs. time during all nights with a calculated ANOVA confidence level  $p < 0.001$ ($>3\sigma$) of variability (a-g). Data points represent individual images obtained with a \emph{Photometrics Star 1} CCD system and Johnson $R$ filter on the 40-cm Newtonian-Cassegrain telescope of FBO. On all figures, flux density (in mJy) is on the y-axis, and time (in UT hours) is on the x-axis. The scale is the same in all figures to better compare among nights. All nights are within 6 days of the peak of the outburst. Error bars are plotted, but some are smaller than the symbols. $F$ and $p$ values are given in Table~\ref{tab:anova}, and $p$ values for an individual night are in the respective figure. (h) June 30 (RJD: 7569.5) is given as a comparison for a non-variable night, with $p = 0.71$.}
\label{fig:varnights}
\end{figure*}

\subsection{Gamma-Ray Outburst}
\label{sec:gammaray}

The \gammaray\ light curve presented in Figure~\ref{fig:gammaonly} is normalized to the maximum flux density of the outburst ($S_\gamma^\mathrm{max} = 22.20 \pm 0.18 \times 10^{-6}$ ph cm$^{-2}$ s$^{-1}$), with time $t=0$ set to the date of maximum ($\sim$ June 24, 19:00, RJD: 7564.3). As mentioned in $\S$\ref{sec:PolObs}, a main flare, $a$, is identified in the outbubrst, along with two smaller-amplitude flares, $b$ and $c$. Flare $b$ occurred $\sim 1$ month after flare $a$, and flare $c$ $\sim 1$ month after flare $b$. All three flares are marked in Figure~\ref{fig:gammaonly}. 

The main flare, $a$, is similar in structure and duration to the optical outburst. Both flux profiles have an asymmetric shape, with the peak occurring late in the flare. The duration of flare $a$ is determined as for the optical outburst: the FWHM of a Gaussian function that fits the flux profile near maximum, $6.3$ days. This duration is close to that of the optical outburst, $\Delta T_\mathrm{opt}^\mathrm{peak} = 5.8$ days. Also, both flare $a$ and the optical outburst exhibit a precipitous decay from peak to pre-outburst levels. Over a similar 24-hour period on June 25 (RJD: 7564.5), flare $a$ decayed by a factor of 10, from $\sim 2.0 \times 10^{-5}$ to $\sim 2.0\times 10^{-6}$ ph cm$^{-2}$ s$^{-1}$. As in the optical case, the decay of flare $a$ occurred mainly over an even shorter timescale, declining by a factor of $\sim 4$ over only 6 hours. 

Unlike the optical outburst, flare $a$ has pre- and post-flare ``plateaus" of enhanced \gammaray\ emission, a common feature of blazar \gammaray\ flares, first discussed by \citet{Abdo2011}. These plateaus, identified visually, are marked by time periods indicated in Figure~\ref{fig:gammaonly}, with durations of $\Delta T_\gamma^\mathrm{pre} \approx 6$ and $\Delta T_\gamma^\mathrm{post} \approx 7.5$ days. In total, the duration of flare $a$ is $\Delta T^a_\gamma = 19.8$ days. Parameters for flare $a$ and the pre- and post-flare plateaus are given in Table~\ref{tab:flareparam}. The timescales of variability are calculated with the same formalism as for the optical light curve. 

The triple-flare structure of a \gammaray\ outburst of 3C454.3 has been seen in previous events, such as the late 2009 (Outburst I), early 2010 (Outburst II), and late 2010 (Outburst III) outbursts analyzed by \citet{Jorstad2013}. Previously, the delay between flare $a$ and $b$ was $\sim 30$ days, and the delay between flares $a$ and $c$ was $\sim 47$ days. For the mid-2016 outburst discussed in this paper, flares $b$ and $c$ occurred later, with delays of $\sim 38$ and $\sim 60$ days from flare $a$, respectively. Another difference between the 2016 outburst and these previous three is the shape of flare $a$. While all four main flares had a pre- and post-flare plateau, the total duration of flare $a$ for the 2016 outburst is less than the duration of the three previous flares by $\sim 10$ days. The extremely short decay of flare $a$ of the 2016 outburst suggests a faster disturbance in our frame or a smaller, more violently variable emission region, as only Outburst III showed a comparable decay range, although over a much longer timescale ($>24$ hours).

Despite the minor differences in shape and timescales, the similarity in structure of the \gammaray\ outbursts argues in favor of a similar mechanism(s) and location of \gammaray\ production for all four events. In fact, the triple-flare structure may be the archetypical pattern of outbursts of 3C454.3. \citet{Jorstad2010} noted a triple-flare structure in the optical light curve of 3C454.3 that coincided in time with the passage of superluminal knots through the mm-wave core of the jet. A similar passage of a knot through sections of the jet containing a relatively high magnetic field and/or relativistic electron density (e.g., a series of standing shocks) could produce the triple-flare structure of the \gammaray\ outbursts. While the measured time interval between the first and third peaks of the earlier events was $\sim 50$ days, it is possible that a similar mechanism/location could result in the structure seen in the 2016 outburst. Parameters of the mid-2016 \gammaray\ outburst, calculated and presented in a manner similar to the outbursts analyzed in \citet{Jorstad2013}, are given in Table~\ref{tab:flareparam}.

\floattable
\begin{deluxetable}{ccccccccc}
\tablecaption{Nights with Significant Variability \label{tab:anova}}
\tablecolumns{2}
\tablewidth{0pt}
\tablehead{\colhead{Night} & \colhead{M} & \colhead{$F$} & \colhead{$p$} & \colhead{$\langle S \rangle$} & \colhead{$\langle \sigma \rangle$} &\colhead{$\Delta T$} & \colhead{$\Delta S$} & \colhead{Sky Conditions} \\
UT Date (RJD) & & & & [mJy] & [mJy] & [hrs] & [mJy]}
\startdata
June 19 (7558.5) & 24 & 29.37 & 6.72$\times 10^{-8}$ &   $7.94 \pm 0.41$ & 0.016 & 2.09 & -0.92  & Full Moon, Clear\\
June 20 (7559.5) & 24 & 10.14 & 1.40$\times 10^{-4}$ &   $9.83 \pm 0.34$ & 0.21  & 1.35 & 0.91 & Full Moon, Local Haze \\
June 22 (7561.5) & 13 & 17.11 & 5.50$\times 10^{-4}$ &  $15.03 \pm 1.09$ & 0.36  & 0.49 & 2.08 & Full Moon, Partial Clouds\\
 & & & & & & & &near End of Night \\ 
June 23 (7562.5)$^a$ & 51 & 113.99&2.94$\times 10^{-23}$& $16.81 \pm 1.52$ & 0.28  & 1.86 & -4.14 & Partial Clouds in Beginning, \\
 & & & & & & & & then Clear after 06:30 UT \\ 
June 24 (7563.5) & 42 & 15.94 & 1.14$\times 10^{-7}$ &  $17.69 \pm 0.73$ & 0.17 & 2.23 & -2.09 & Clear \\
June 25 (7564.5) & 47 & 178.43& 9.15$\times 10^{-26}$& $ 7.72 \pm  1.33$ & 0.16 &2.90 & -3.69 & Clear\\
June 26 (7565.5) & 40 & 16.61 & 1.38$\times 10^{-7}$ &  $ 4.20 \pm 0.43 $ & 0.21 & 2.05 & -1.19 & Clear \\
June 30 (7569.5) & 48 & 0.53  &  0.71                         &  $ 2.28 \pm 0.14$ & 0.14 & 0.55 & -0.42 & Clear \\
\enddata
\tablecomments{$^a$Due to unfavorable weather conditions, 20 of the first 22 images during the night were 60-second exposures. The parameters are labeled as follows: $M$: Number of observations during the night; $F$: $F$-value calculated from ANOVA test; $p$: $p$-value calculated from ANOVA test, interpreted as significantly variable if $p \leq 0.001$ ($>3\sigma$); $\langle S \rangle$: Average flux-density during the night, with 1$\sigma$ standard deviation; $\langle \sigma \rangle$: Average error per measurement; $\Delta T$: Time between third-highest and third-lowest flux density levels, in hours; $\Delta S$: Flux density difference between the third-highest and third-lowest flux densities. Negative values indicate that the flux density decreased over the course of the night.}
\end{deluxetable}

%------------------------------------------------------------------------------------------------------------------------
% 		MICRO-VARIABILITY
%------------------------------------------------------------------------------------------------------------------------

\section{Rapid Optical Flux Variability}
\label{sec:variability}

3C454.3 had been observed to have variations in brightness only on the order of 1 magnitude over the course of entire observing seasons \citep{Angione1968, Lloyd1984, Webb1988} until several outbursts during the late 2000s, most notably the unprecedented 2005 outburst \citep{Villata2006}, with a peak brightness of $R = 12$. Inspection of the optical $R$ band light curve during the 2016 outburst reveals several periods of intense variability over the course of a single night. We refer to such events as ``intraday variability" if the light curve appears to connect smoothly with the flux on the preceding and subsequent nights. The term ``micro-variability" is reserved to describe changes during a night when the behavior of the flux deviates from the interday variability trend. This section describes observations of notable intraday variability, as well as one night with clearly evident micro-variability in the form of quasi-periodic oscillation of the optical flux. Only data obtained from FBO are used in this analysis, since none of the other optical telescopes involved in this study observed 3C454.3 continuously over a given night.

\subsection{Intraday Variability}
\label{sec:intraday}

In order to increase the confidence of reports of variability in light curves, several statistical methods have been developed to quantify the variations of sources. For example, \citet{DeDiego2010} has provided a direct comparison of several statistical tests and determined that a one-way analysis of variance (ANOVA) test is a robust method to detect and quantify variations from AGN. Applied to quasar variability, an ANOVA test checks for the probability of several sample groups being equal, with the null hypothesis for an ANOVA test being non-variability. We utilized a standard ANOVA test instead of a more complicated enhanced $F$-test or Bartels test \citep{DeDiego2014}, as only a single non-variable bright comparison star was used during the photometry, and the light curve on each night is oversampled compared to the timescale of variations being determined. More robust statistical methods can be used \citep[e.g.,][]{DeDiego2015}, but in our case an ANOVA test is sufficient. 

In this analysis, an ANOVA test was first used to eliminate potential variability of comparison stars in the same field as 3C454.3. The ANOVA test was then applied to each night of data of 3C454.3 collected from FBO. The ANOVA test revealed 8 nights of possible intraday variability at the $p < 0.001$ confidence level ($>3\sigma$), on every night of observations between June 19 and June 26 (RJD: 7558.5-7565.5), plus July 21 (RJD: 7590.5). However, since July 21 was after the outburst, we ignore that night's data for the rest of the analysis. On several other nights, the flux varied at a confidence level $<3\sigma$; we also ignore the data from these nights. 

Light curves for nights between June 19 and June 26 (RJD: 7558.5-7565.5) are given in Figure~\ref{fig:varnights}. June 30 (RJD: 7569.5) is also included as a control, since no variability of 3C454.3 was observed on that night according to the ANOVA test. Important calculated values for the data on each night are presented in Table~\ref{tab:anova}, as is the sky condition. The calculated $F$ and $p$ values from each ANOVA test are given in Table~\ref{tab:anova}, and the $p$ values are also included in Figure~\ref{fig:varnights}. Smaller $p$ values indicate a higher probability of the source being variable during the night. The average flux density, $\langle S \rangle$ (with 1$\sigma$ standard deviations), the timescale of variability, $\Delta T$, and the change in flux density of the source, $\Delta S$, were calculated for each night. The change in flux during each night and the time difference between maximum and minimum points were calculated using the third-highest and third-lowest flux density values to avoid the influence of outlying data points in the analysis. 

The most significant variability was seen on June 23 and 25 (RJD: 7562.5 and 7564.5).  The total change in flux density of 3C454.3 during these two nights was $\Delta S = - 4.14$ mJy and $\Delta S = -3.69$ mJy, where the negative value of $\Delta S$ indicates a decrease of flux density during the night. 

\subsection{Micro-Variability}
\label{sec:Micro}

The light curve of 3C454.3 on June 25 (RJD: 7564.5) exhibits significant intraday variability (see Fig.~\ref{fig:varnights} (f)), with the flux density decreasing by a factor $\sim2$ over the 3.5 hours of observation. This is the steepest observed rise or fall in optical flux density throughout the outburst. Since the shortest timescales of variability are important for constraining the physical size of the emission region, the variability of 3C454.3 on June 25 is now examined in detail. 

\begin{figure}[t!]
\plotone{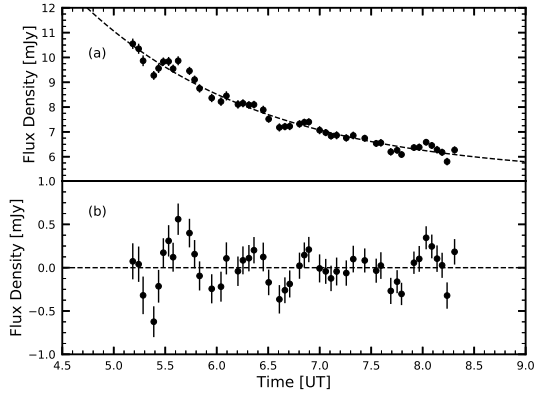}
\caption{$(a)$ Intraday variability of 3C454.3 on June 25 (UT). The data are parameterized by a single decaying exponential (dashed line). $(b)$ Residuals of the exponential fit. In both panels, error bars are included.}
\label{fig:June25expo}
\end{figure}

Following \citet{Valtaoja1999}, we roughly model the optical light curve of 3C454.3 on June 25 with an exponential rise and decay of the form $S = A \exp{(-Bt)} + C$. It is only possible to obtain a rough estimation for the value of $A$ ($\sim100$ mJy), since the amplitude of the exponential is changing drastically over a short period of time. This large value is a consequence of fitting such a small section of the data and is not meant to connect the beginning of the light curve on June 25 with the end of the light curve on the preceding night. The fitted parameters $B = 0.569 \pm 0.004$ hr$^{-1}$ and $C = 5.20 \pm 0.08$ mJy are more robust. Application of this fit and the resulting residuals are shown in Figure~\ref{fig:June25expo}. 

The residuals of the fit show oscillations above and below the exponential trend throughout the night. These oscillations are more pronounced than short-term scatter. We have checked that observational effects, such as weather, atmospheric reddening, or placement of the source on the CCD had no impact on the observed flux density. The oscillations appear to be largely independent of the aperture used during photometry. Thus, we judge these variations to be intrinsic to the blazar. 

In order to model these oscillations, we use a decaying sinusoid of the form

\begin{equation*}
S = \frac{f(t)}{f_0} (A_0 \sin{(B_0 t-C_0)})\  ,
\end{equation*}

\noindent where $f(t)$ is the flux density of the exponential trend, $f_0$ is a reference value chosen to be 6 mJy (the flux density of the blazar at $\sim 8.5$ UT), and $A_0, B_0,$ and $C_0$ are constants. A decaying sinusoid was adopted to better represent the data at the beginning and end of the night, since the flux density oscillation amplitude decreased during the night. 

\begin{figure}
\plotone{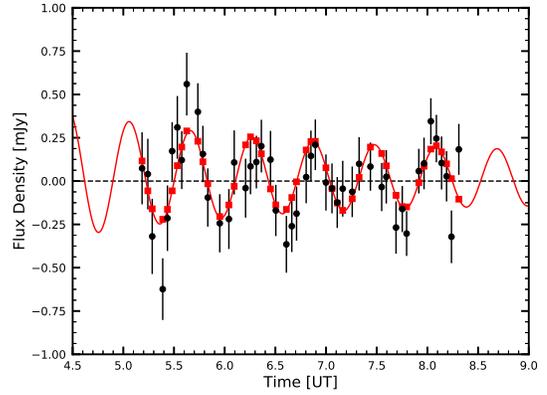}
\caption{Sinusoidal fit (red solid line) to the residuals of the exponential decay on June 25 (black circles). See text for details.}
\label{fig:sinonly}
\end{figure}

The sinusoidal fit to the residual data is shown in Figure~\ref{fig:sinonly}. The constants calculated from the fit are $A_0 = 0.17 \pm 0.01$ mJy, $B_0 = 10.39 \pm 0.03$ hr$^{-1}$  (period $= 36.28 \pm 0.09$ minutes), and a phase factor $C_0 = 0.750 \pm 0.002$. There are insufficient data to determine whether the oscillation pattern lined up with any data a few hours before or after the FBO observations.  Although a changing frequency of oscillation may more accurately fit the observed oscillations, there are too few data points to justify such a complication.

\begin{figure*}
\plotone{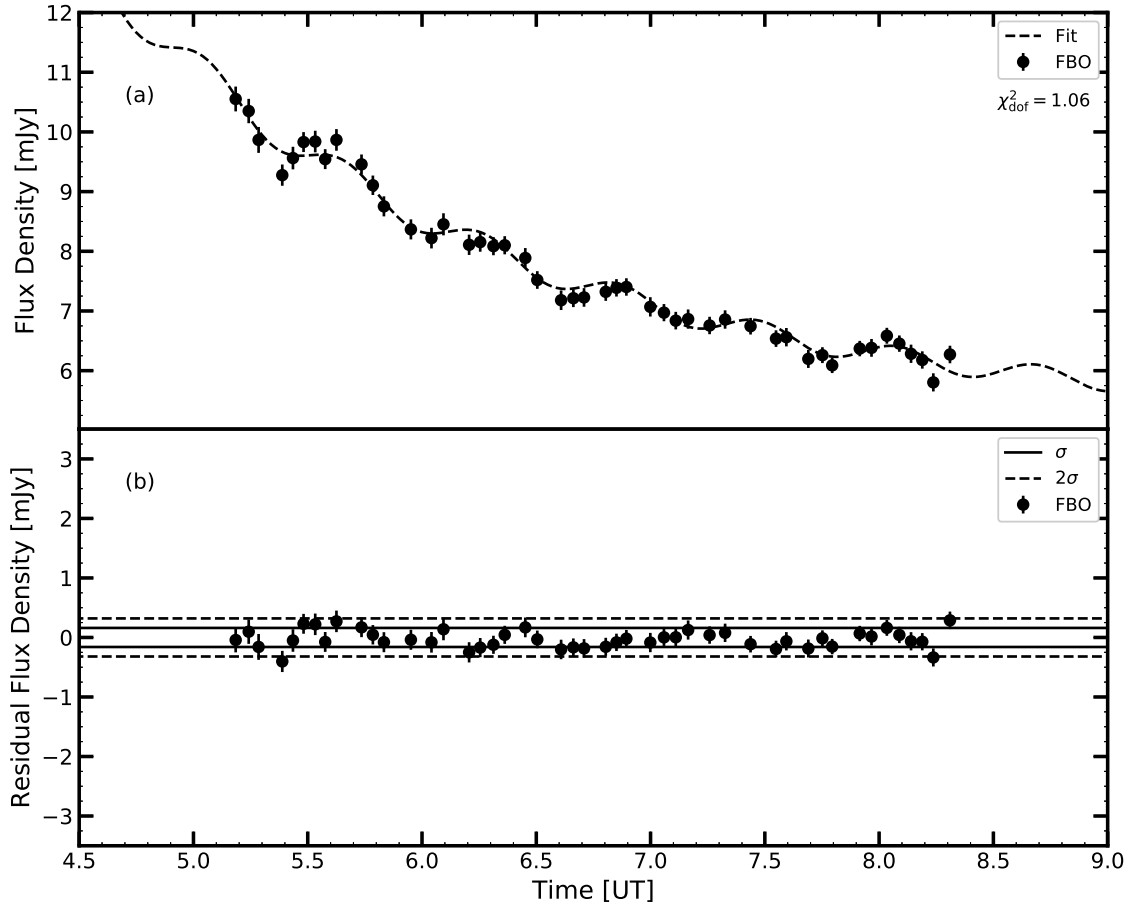}
\caption{$(a)$ Exponential and sinusoidal fit combined together to model the behavior of 3C454.3 on June 25 (RJD: 7564.5). See text for details. $(b)$ The residual flux from the fit, on the same scale as the light curve. The solid and dashed black lines in $(b)$ represent the average $1\sigma$ and $2\sigma$ error bars. A $\chi^2$ test for goodness of fit yields $\chi_\mathrm{dof}^2 = 1.06$ for the fit. \label{fig:finalfit}}
\end{figure*}

Combining both the exponential and sinusoidal models produces a fit that yields very low residual flux density levels, as seen in Figure~\ref{fig:finalfit}. The residual scatter fits mostly within two standard deviations of the average error bar. The chi-square per degree of freedom of the total combined exponential and sinusoid model $\chi_{\mathrm{dof}}^2 = 1.06$. 

The micro-variability oscillations on June 25 show that 3C454.3 can vary significantly on sub-hour timescales, in this case 36 minutes. This severely constrains the size of the emission region as discussed in $\S$\ref{sec:discuss}.

%------------------------------------------------------------------------------------------------------------------------
% 		DISCUSSION
%------------------------------------------------------------------------------------------------------------------------

\section{Discussion}
\label{sec:discuss}

The multi-frequency light curves reveal the extraordinary June 2016 outburst of 3C454.3, with rapid, high-amplitude changes in flux density over short timescales. At \gammaray\ energies, 3C454.3 varied on timescales at least as short as $\sim 3$ hours. The necessity of binning data prevents us from detecting any more rapid variations that may have occurred. At optical $R$ band, we observed flux density variations of more than 1 mJy per hour near the peak of the outburst, with a minimum timescale of variability $\tau_\mathrm{opt}^\mathrm{min} \approx 2$ hours. Observations on June 25 (RJD: 7564.5) revealed micro-variability in the form of quasi-periodic oscillations with an estimated period of 36 minutes.

%------------------------------------------------------------------------------------------------------------------------
% 		JET VARIABILITY
%------------------------------------------------------------------------------------------------------------------------

\subsection{Overall Variability in the Jet}
\label{sec:overallvariability}

It is possible to relate the observed timescale of variability $\tau_\mathrm{var}(\mathrm{obs})$ to the intrinsic value $\tau_\mathrm{min}^\mathrm{intr}$ in the rest frame of the blazar using

\begin{equation*}
\tau_\mathrm{min}^\mathrm{intr} = \frac{\delta \tau_\mathrm{var}(\mathrm{obs})}{1+z}\ ,
\end{equation*}

\noindent where $z$ is the redshift of the host galaxy and $\delta$ is the Doppler factor. A technique developed by \citet{Jorstad2005} derives the Doppler factor $\delta$ through analysis of 43 GHz Very Long Baseline Array (VLBA) images. The movement of bright ``knots" down the jet often coincides with flux outbursts. Knots can have different speeds and values of $\delta$ \citep[e.g.,][]{Jorstad2001, Jorstad2010, Kellermann2004, Lister2009}. 

%------------------------------------------------------------------------------------------------------------------------
% 		RADIO KNOTS
%------------------------------------------------------------------------------------------------------------------------

\subsubsection{Radio Knots}
\label{sec:RadioKnots}

\begin{figure}
\epsscale{0.775}
\plotone{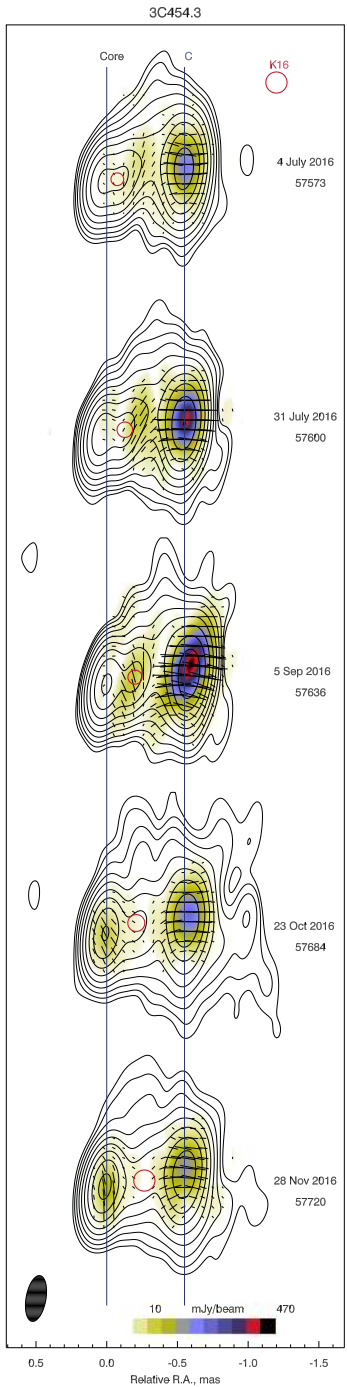}
\caption{VLBA total (contours) and polarized (color scale) intensity images of 3C454.3 at 43 GHz showing the evolution of $K16$, convolved with a beam of $0.33 \times 0.14$ mas$^{2}$ at $\mathrm{PA} = -10\degr$ (the bottom left grey-lined oval). The global intensity peak is $7300$ mJy/beam, and contour levels start at 0.1\% of the peak and increase by a factor of 2. Black line segments within each image show the direction of linear polarization, while the length of the segment is proportional to the polarized intensity values. The black and navy vertical lines indicate the position of the core and stationary feature $C$ \citep[see][]{Jorstad2017}, respectively, while the red circles indicate the position and size of $K16$ according to modeling. \label{fig:K16}}
\end{figure}

\begin{figure*}
\plotone{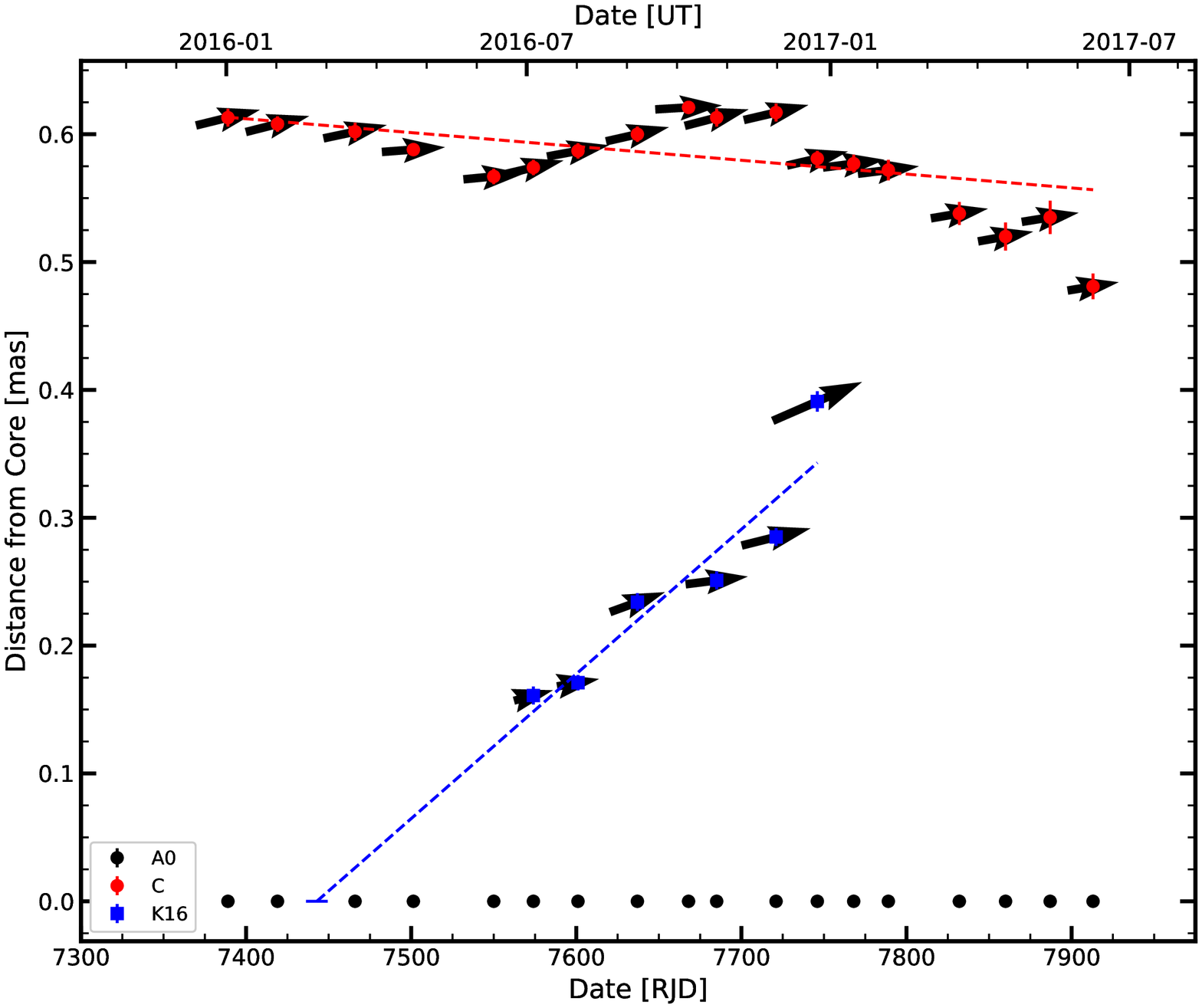}
\caption{Separation of $K16$ from the core vs. time in the jet of 3C454.3 from the VLBA-BU-BLAZAR monitoring program. The vectors show the position angle of each knot with respect to the core at the corresponding epoch. The dashed lines represent polynomial fits to the motion, as done in \citet{Jorstad2017}. The black dots mark the position of the core, $A0$. The red points correspond to the stationary feature $C$, while the blue points correspond to $K16$. The vertical line segments show the approximate $1\sigma$ positional uncertainties based on the brightness temperature $T_\mathrm{b}$. The horizontal line segment for $K16$ shows uncertainty in the date of ejection of the knot from the core. \label{fig:Distance}}
\end{figure*}

We have analyzed the VLBA data obtained at 43 GHz within the VLBA-BU-BLAZAR program\footnote{\url{http://www.bu.edu/blazars/VLBAproject.html}} from 2016 January to 2017 June. The data reduction and model fitting are described in \citet{Jorstad2017}. The analysis reveals a knot ejected from the 43 GHz core of the blazar near in time to the 2016 outburst, designated $K16$. The knot was first distinguishable from the 43 GHz core in June 2016, coincident in time with the optical and \gammaray\ outburst. Figure~\ref{fig:K16} presents a sequence of 43 GHZ total and polarized intensity VLBA images of 3C454.3 depicting the evolution of $K16$. Following $K16$ until it was no longer visible, the knot had an apparent speed of $v_\mathrm{app} = 20.3c \pm 0.8c$. A backward extrapolation of the motion under the assumption of constant speed yields a date of 2016 February 25 (RJD: 7443 $\pm$ 17.5 days) when the brightness centroid of the knot crossed that of the core. Using the method described in \citet{Jorstad2017}, the Doppler factor $\delta = 22.6 \pm 4.4$, bulk Lorentz factor $\Gamma = 20.4 \pm 0.4$, and viewing angle of the path of $K16$ with respect to the line of sight $\Theta_\circ = 2.5 \degr \pm 0.3\degr$. Knot $K16$ has a wider viewing angle than $K09$ and $K10$ (associated with Outbursts I-III, $\S$\ref{sec:gammaray}), that moved down the jet along paths oriented $1.35 \degr \pm 0.2\degr$ and $0.4\degr \pm 0.1\degr$ from our line of sight \citep{Jorstad2013}. Figure~\ref{fig:Distance} shows the separation of $K16$ from the core, in addition to the stationary feature $C$ located $\sim0.58$ mas from the core.

$K16$ was ejected from the core $\sim \mathbf{4}$ months before the outburst. The time delay could be shorter if the knot decelerated as it separated from the core. The angular sizes of K16 and the core when the VLBA observations could first resolve them separately were $0.2 \pm 0.02 $ mas and $0.1 \pm 0.02 $ mas, respectively. Since in 4 months $K16$ moved $\sim 0.1$ mas, the upstream boundary of the knot was still crossing the core when the outburst occurred. In multiwavelength outbursts of other sources \citep[such as the BL Lacertae object AO 0235+164, see][]{Agudo2011}, superluminal knots have been seen as the ``head" of an extended disturbance containing a front-back structure stretched by light-travel delays in the observer's frame \citep[e.g.,][]{Aloy2003}. Then, when the back perturbation encounters the core, particle acceleration causes the observed muliwavelength variability. The timing of $K16$ and  the multiwavelength variability observed is consistent with a lagging upstream end of $K16$ causing the outburst as it interacted with a standing shock in the core. 

\begin{figure}
\plotone{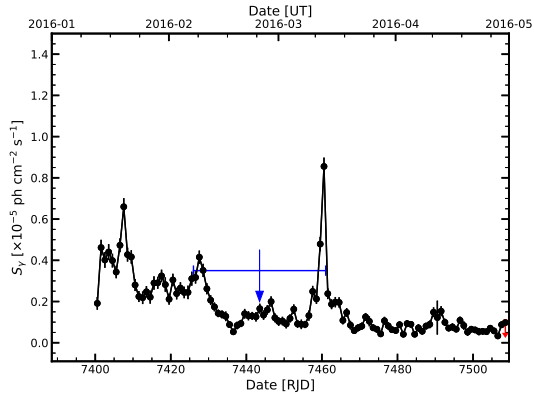}
\caption{\fermi-LAT \gammaray\ flux of 3C454.3 with daily time bins prior to the 2016 outburst. Upper limits are marked with a downward facing, red arrow. The downward facing,  blue arrow marks the date of ejection of $K16$: 2016 February 25 (RJD: 7443 $\pm$ 17.5 days).}
\label{fig:ExtraGamma}
\end{figure}

Analysis of \gammaray\ data collected with the \fermi-LAT in the months preceding the June 2016 outburst reveals a small amplitude \gammaray\ outburst with $S_\gamma^\mathrm{max} = 8.55\pm0.42 \times 10^{-6}$ ph cm$^{-2}$ s$^{-1}$ on $T_\gamma^\mathrm{max} = $  2016 March 13 (RJD: 7460, see Figure~\ref{fig:ExtraGamma}). The timing of this flare is consistent with the event being caused by a forward section of $K16$ passing through the core. 

The position angle of the jet projected on the sky for $K16$ and the stationary component $C$ during the 2016 outburst was $-79\degr \pm 1 \degr$. This position angle is outside the 5-year average found in \citet{Jorstad2017} of $-98\degr \pm 10\degr$. However, the jet position angle during the 2016 outburst is consistent with the average position angle of the knot $B10$ \citep{Jorstad2017}. In the range of polarization angles presented in Figure~\ref{fig:lightcurves} (d), values $\chi_\text{opt} \sim 101\degr$ are parallel to the jet direction, while either $\chi_\text{opt} \sim 11\degr$ or $191\degr$ are nearly perpendicular to the jet direction.\\

$\left. \right.$

%------------------------------------------------------------------------------------------------------------------------
% 		Magnetic Field Strength
%------------------------------------------------------------------------------------------------------------------------

\subsubsection{Magnetic Field Strength}
\label{sec:MagFieldStrength}

For this analysis, we use the Doppler factor of $K16$, $\delta = 22.6$. With $z = 0.859$ for 3C454.3, the timescale of variability in the rest frame of the emitting plasma in $R$ band is $\tau_\mathrm{min}^\mathrm{intr} \approx 24$ hours. The maximum size of the emission region is related to the intrinsic variability timescale through relativistic causality: $r \lesssim c\tau_\mathrm{min}^\mathrm{intr} \approx 2.6\times 10^{15}$ cm.

The intrinsic timescale of variability can also be used to provide an estimate for the strength of the magnetic field in the jet. For shock-in-jet models of blazar variability \citep[e.g.,][]{Marscher1985}, the shock energizes relativistic electrons as they enter the emitting region behind the shock front. Both synchrotron and inverse Compton radiative losses then determine the extent of the emission region in the direction of the jet flow. If the spectral energy distribution at infrared-optical and \gammaray\ frequencies is similar to that of the 2010 November outburst \citep[Fig. 25 of][]{Jorstad2013}, the ratio of inverse Compton (\gammaray) to synchrotron (IR) luminosity at the peak of the 2016 June outburst is $\sim 5$. (The 2010 November outburst spectral energy distribution is used due to the lack of available X-ray, UV, and infrared data for the 2016 outburst, thus the peaks in the spectral energy distribution cannot be determined.) In the electron energy loss equation, the total loss rate is derived in part by summing both the energy density due to the relativistic particles and the magnetic field. Since these quantities are proportional to the inverse Compton and synchrotron peaks respectively, if the ratio of the inverse Compton to synchrotron luminosity is $5:1$, then the expression $u_\mathrm{ph} + B^2 / 8\pi$ can be reduced to $6B^2 / 8\pi$. In the observer's frame, the lifetime of electrons emitting at a frequency $\nu_\mathrm{GHz}$ (in GHz) can be related to the magnetic field strength $B_\mathrm{G}$ (in Gauss) through

\begin{equation*}
B_\mathrm{G} \approx \left(\frac{1+z}{6 \delta \nu_\mathrm{GHz}} \left(\frac{4.75\times10^2}{t_\mathrm{loss,days}} \right)^2 \right)^\frac{1}{3}\ ,
\end{equation*}

\noindent where $t_\mathrm{loss,days}$ is the timescale of energy loss in days \citep[e.g.,][]{Hagenthorn2008}. At $R$ band (central frequency of $\nu_\mathrm{GHz} = 4.69\times 10^5$ GHz), the observed value of $t_\mathrm{loss,days} = \tau_\mathrm{opt,days}^\mathrm{min} = 0.082$ yields $B_\mathrm{G} \approx 1$ Gauss during the outburst. This magnetic field strength is a factor of $\sim2$ higher than other estimates in the optically emitting region of a blazar (e.g., \citet{Hagenthorn2008}), as expected during an outburst.

%------------------------------------------------------------------------------------------------------------------------
% 		JET MICROVARIABILITY
%------------------------------------------------------------------------------------------------------------------------

\subsection{Micro-Variability in the Jet}
\label{sec:shortvariability}

Micro-variability, as defined in $\S$\ref{sec:variability}, was observed on 2016 June 25 (RJD: 7564.5) in the $R$ band light curve of 3C454.3. It can be described as quasi-periodic oscillations, with an amplitude of $2 - 3\%$ (corresponding to 0.17 mJy) about an exponentially decreasing trend (from 11 to 6 mJy) over 3.5 hours, with an oscillation period of 36 minutes. 

The micro-variability cannot be explained by an emission region as large as that derived in $\S$\ref{sec:overallvariability}, $r \lesssim 2.6\times10^{15}$ cm, without violating causality unless one invokes a contrived geometry \citep[see][]{Spada1999}. Instead, for the observed timescale of micro-variability, the intrinsic timescale is $\tau_\mathrm{min}^\mathrm{intr} \approx 7$ hours. The size of an emission region capable of varying on such a timescale is $\lesssim 8 \times 10^{14}$ cm. 

Statistical analyses of the flux variations in blazars have shown that the variations are governed by noise processes with higher amplitudes on longer timescales \citep[e.g.,][]{Chatterjee2008, Abdo2011}.  The presence of a helical magnetic field \citep{Lyutikov2005, Pushkarev2005} can explain the range of the degree and position angle of polarization measured in 3C454.3 during the 2016 outburst \citep[and in blazars in general, e.g.,][]{Jorstad2007}. However, the rapid, seemingly random variations observed (see $\S$\ref{sec:opticalpolarization}) are naturally reproduced in neither a 100\% globally ordered nor completely chaotic (on small scales) field. Instead, a more natural explanation for the fluctuations of flux and polarization is the presence of turbulent plasma in the relativistic jets of blazars. This type of emission has been the subject of various simulation studies of blazar variability \citep[e.g.,][]{Marscher2014, Calafut2015, Pollack2016}. If one approximates the pattern of turbulence in terms of $N$ turbulent cells, each with a uniform magnetic field with random orientation, then the degree of linear polarization has an average value of $\langle \Pi \rangle \approx \Pi_\mathrm{max} N^{-1/2}$, where $\Pi_\mathrm{max}$ corresponds to a uniform field case and is typically between 0.7 and 0.8 \citep{Burn1966}. The polarization will vary about the mean with a standard deviation $\sigma_\Pi \approx 0.5 \Pi_\mathrm{max} N^{-1/2}$ if the cells pass into and out of the emission region \citep{Jones1988}. 

We adopt this model to explain the variation of $P_\mathrm{opt}$ and $\chi_\mathrm{opt}$ over time. The average degree of polarization during the outburst in 3C454.3 (2016 June 10-28, RJD: 7549.5-7567.5) was $\langle \Pi \rangle = 12.5\% \pm 5.4\%$. For the median value $\Pi_\mathrm{max} = 75\%$, the number of turbulent cells during the outburst was $N = (\Pi_\mathrm{max}/\langle \Pi\rangle)^2 \approx (0.75/0.125)^2 \approx 35$ cells. The expected standard deviation of the fluctuations of $P_\mathrm{opt}$ is then $\sigma_\Pi = 6.3\%$, which is similar to the observed standard deviation, 5.4\%. 

The behavior of $\chi_\mathrm{opt}$ in comparison with this model can provide new insight into the mechanics of the turbulent cells. All cells in the model have their own uniform magnetic field with a random orientation. Prior to the 2016 outburst, a low level of polarization of 3C454.3 was measured (see Fig.~\ref{fig:lightcurves}c), which requires a large number ($\gtrsim 100$) of cells. Only $\sim 35$ cells participated in the outburst, perhaps via a number of magnetic reconnections of the turbulent magnetic field that rapidly accelerated electrons to energies $\gtrsim 10^4$ mc$^2$ \citep[e.g.][]{Kadowaki2015}. The turbulence could have been enhanced by feedback between the reconnections and chaotic motions of the plasma \citep{Lazarian2016}. This might have led to clustering of the cells that most efficiently accelerated electrons to cause a more coherent flux outburst.

The size of the emission region, $r \lesssim 8 \times 10^{14}$, involving $\sim35$ turbulent cells gives the size of each turbulent cell as $r_\mathrm{cell} \lesssim 2.3 \times 10^{13}$ cm. \citet{Calafut2015} simulated blazar light curves under a variety of conditions with a rotating turbulent cell model for the jet. The turbulent cells were considered roughly spherical, and differential Doppler beaming the eddies could be responsible for variations in the light curves of blazars ranging from a few percent to large, chaotic outbursts, depending on the speed of the turbulent motions. From the simulations, \citet{Calafut2015} determined that, in order to generate simulated light curves similar to those observed, the rotation speed of the turbulence should be $0.1c \leq v_\mathrm{cells} \leq 0.3c$, and that higher turbulent velocities should not be common. 

The estimated size of the cells can be combined with the period of the quasi-periodic micro-variability found in $\S$\ref{sec:Micro} to provide an observational measurement of the speed of the turbulent motions in a blazar jet. If we approximate that the turbulent cells are rotating cylinders, the period of the oscillations, $P$, can be related to the angular speed, $\omega$, of rotation through $\omega = \frac{2\pi}{P}$. In the rest frame of the blazar, the period of oscillations is $\sim 7$ hours, which yields $\omega \approx 2\times 10^{-4}$ rad s$^{-1}$. The tangential velocity $v_\mathrm{cell} = \omega r_\mathrm{cell} \simeq 0.2c$. This value of the turbulent speed agrees with results of the simulations in \citet{Calafut2015}, as does the $2-3\%$ level of the quasi-periodic oscillations in flux observed in 3C454.3. 

%------------------------------------------------------------------------------------------------------------------------
% 		CONCLUSIONS
%------------------------------------------------------------------------------------------------------------------------

\section{Conclusions}
\label{sec:conclusions}

At both optical and \gammaray\ frequencies, the flux density from 3C454.3 increased over a 1.5-week rising time and featured a precipitous decay over the course of $\sim 24$ hours. The peaks of the \gammaray\ and optical light curves are coincident to within a 24-hour binning of the \gammaray\ light curve, suggesting that the location of the enhanced \gammaray\ and optical emission is similar. Prior to the outburst, the $R$ band degree of polarization $P_\mathrm{opt}$ was low, and the angle of polarization $\chi_\mathrm{opt}$ was roughly parallel to the radio jet axis. During the outburst, $P_\mathrm{opt}$ changed to $\sim 20\%$, with $\chi_\mathrm{opt}$ changing in an irregular fashion by $\sim 120\degr$. In general, the optical spectropolarimetric observations obtained during the outburst indicate that $P_\mathrm{opt}$ is dependent on wavelength, decreasing towards the blue end of the spectrum, which can be attributed to dilution of the polarization from the unpolarized blue bump emission. The polarization decreases at the Mg II line, indicating that the broad line region does not have a strong polarization. The high time-resolution light curve on 2016 June 25 (RJD:7564.5) reveals micro-variability in the form of quasi-periodic oscillations with an amplitude of $2$-$3\%$ around the mean trend and a period of 36 minutes.

Analysis of 43 GHz VLBA maps of the total and polarized intensity of 3C454.3 indicates a ``knot" of plasma, $K16$, ejected from the radio core of the blazar near in time to the 2016 outburst on 2016 Feb 25 (RJD: 7443 $\pm$ 17.5). This knot is likely responsible for a small amplitude \gammaray\  outburst in 2016 March when the ``head" of the knot moved past the standing shock, as well as the main 2016 outburst when the lagging end moved past the shock.

From the time dependence of the optical $R$ band flux density, polarization degree, and position angle, the following physical characteristics of the jet of 3C454.3 can be determined. The minimum observed timescale of variability $\tau_\mathrm{opt}^\mathrm{min} \approx 2$ hours. The intrinsic timescale of variability in the rest frame of the emitting plasma $\tau_\mathrm{min}^\mathrm{intr} \approx 24$ hours, based on a Doppler factor $\delta = 22.6$ from the VLBA analysis. Relativistic causality restricts the size of the emission region to $r \lesssim 2.6\times 10^{15}$ cm. If the timescale of flux decline corresponds to the energy loss time of the radiating electrons, the magnetic field in the jet $B_G \approx 1$ Gauss, under the assumption that the ratio of the inverse Compton to synchrotron luminosity is 5:1, as during the November 2010 outburst.

A shock-in-jet model with turbulence can naturally explain the observed variability in the light curves. From the micro-variability oscillations, the average optical degree of polarization and its variations during the outburst, we estimate the size of a single turbulent cell to be $r_\mathrm{cell} \lesssim 2.3 \times 10^{13}$ cm. The speed of rotation of the turbulent cell is then $\sim 0.2c$. This value is in agreement with simulations of the effect of turbulence on blazar light curves \citep{Calafut2015}, which predict a change in flux density on the order of a few percent for a turbulent speed $0.1c \leq v \leq 0.3c$. The turbulence could have been responsible for the outburst through a series of magnetic reconnection events that rapidly accelerated electrons to energies capable of producing optical synchrotron and \gammaray\ inverse Compton photons.

%------------------------------------------------------------------------------------------------------------------------
% 		ACKNOWLEDGEMENTS
%------------------------------------------------------------------------------------------------------------------------

\acknowledgements ZW gratefully acknowledges support through Colgate University's Justus and Jayne Schlichting Student Research and the Division of Natural Sciences and Mathematics funds. The knowledge of Prof. Sara Buson, Dr. Elizabeth Ferrara, and the rest of the Fermi Team was invaluable in helping to perform the \gammaray\ analysis. The assistance of the other undergraduate observers at Colgate University during the observing campaign is greatly appreciated. The research at Boston University was supported in part by National Science Foundation grant AST-1615796 and NASA Fermi Guest Investigator grant 80NSSC17K0649. Data from the Steward Observatory spectropolarimetric monitoring program were used. This program is supported by Fermi Guest Investigator grant NNX15AU81G. The St. Petersburg University team acknowledges support from Russian Science Foundation grant 17-12-01029. The VLBA is an instrument of the National Radio Astronomy Observatory. The National Radio Astronomy Observatory is a facility of the National Science Foundation operated under cooperative agreement by Associated Universities, Inc. We gratefully thank the anonymous referee for comments and useful suggestions that helped to improve this work. This research made use of Astropy, a community-developed core Python package for Astronomy \citep{Astropy2013, Astropy2018}.

\facilities{Perkins (PRISM),  CrAO: 0.7m AZT-8, SPbU: 0.4m LX-200, SO:Bok, Kuiper (SPOL), Fermi (LAT), Foggy Bottom Observatory, VLBA}

\software{IRAF, \fermi\ Science Tools, HEASoft, Python, Astropy}

{}

\end{document}